**JSWSC**

RESEARCH ARTICLE                                                    OPEN ⭗ ACCESS

# A homogeneous *aa* index: 1. Secular variation

Mike Lockwood[1,*], Aude Chambodut[2], Luke A. Barnard[1], Mathew J. Owens[1], Ellen Clarke[3],
and Véronique Mendel[4]

[1]  Department of Meteorology, University of Reading, Whiteknights Campus Earley Gate, PO Box 243, Reading RG6 6BB, UK
[2]  Institut de Physique du Globe de Strasbourg, UMR7516; Université de Strasbourg/EOST, CNRS, 5 rue René Descartes,
    67084 Strasbourg Cedex, France
[3]  British Geological Survey, Edinburgh EH14 4AP, UK
[4]  International Service of Geomagnetic Indices, 5 rue René Descartes, 67084 Strasbourg cedex, France

**Abstract** – Originally compiled for 1868–1967 and subsequently continued so that it now covers
150 years, the *aa* index has become a vital resource for studying space climate change. However, there
have been debates about the inter-calibration of data from the different stations. In addition, the effects
of secular change in the geomagnetic field have not previously been allowed for. As a result, the compo-
nents of the "classical" *aa* index for the southern and northern hemispheres ($aa_S$ and $aa_N$) have drifted
apart. We here separately correct both $aa_S$ and $aa_N$ for both these effects using the same method as used to
generate the classic *aa* values but allowing $\delta$, the minimum angular separation of each station from a nom-
inal auroral oval, to vary as calculated using the IGRF-12 and gufm1 models of the intrinsic geomagnetic
field. Our approach is to correct the quantized $a_K$-values for each station, originally scaled on the assump-
tion that $\delta$ values are constant, with time-dependent scale factors that allow for the drift in $\delta$. This requires
revisiting the intercalibration of successive stations used in making the $aa_S$ and $aa_N$ composites. These
intercalibrations are defined using independent data and daily averages from 11 years before and after
each station change and it is shown that they depend on the time of year. This procedure produces
new homogenized hemispheric *aa* indices, $aa_{HS}$ and $aa_{HN}$, which show centennial-scale changes that
are in very close agreement. Calibration problems with the classic *aa* index are shown to have arisen from
drifts in $\delta$ combined with simpler corrections which gave an incorrect temporal variation and underesti-
mate the rise in *aa* during the 20th century by about 15%.

## 1 Introduction

### 1.1 The derivation of the classic *aa* index

In his book (Mayaud, 1980), Pierre-Noël Mayaud attributes
the origins of the idea for the *aa* index to the 1969 IAGA
(International Association of Geomagnetism and Aeronomy)
meeting in Madrid, where a request for an effort to extend geo-
magnetic activity indices back in time was made by Sydney
Chapman on behalf of the Royal Society of London. Mayaud's
subsequent work resulted in an index somewhat different from
that which Chapman had envisaged, but which covered
100 years between 1868 and 1967 (Mayaud, 1971) and has
become a key component of research into space climate
change. This index, termed *aa*, was adopted at the 1975 IAGA

meeting in Grenoble (IAGA, 1975). It was made possible by
the availability of magnetic records from two old observatories,
Greenwich in southern England and Melbourne in Australia.
These two stations are almost antipodal, roughly at the same
geomagnetic latitude and 10 h apart in local time. To make a
full data sequence that extends from 1868 to the present day,
it is necessary to use 3 stations in each hemisphere. In England
they are: Greenwich (IAGA code GRW, 1868–1925), Abinger
(ABN, 1926–1956, 51.185°N, 359.613°E), and Hartland (HAD,
1957–present, 50.995°N, 355.516°E). In Australia they are:
Melbourne (MEL, 1868–1919, −37.830°N, 144.975°E),
Toolangi (TOO, 1920–1979, −37.533°N, 145.467°E) and
Canberra (CNB, 1980–present, −35.315°N, 149.363°E).
The *aa* index is based on the *K* values for each station, as
introduced by Bartels et al. (1939). These are derived from the
range of variation observed at the station in 3-hour intervals.

*Corresponding author: m.lockwood@reading.ac.uk

The formal procedure for deriving $K$ is: the range (between minimum and maximum) of the irregular variations (that is, after elimination of the regular daily variation) observed over a 3-hour interval in either of the horizontal components ($X$ northward or $Y$ eastward, whichever gives the larger value) is ranked into 1 of 10 classes (using quasi-logarithmic band limits that are specific to the observatory) to which a $K$ value of 0–9 is assigned. The advantage of this procedure is that the scale of threshold values used to convert the continuous range values into the quantized $K$ values is adjusted for each station to allow for its location and characteristics such that the $K$ value is a standardized measure of the geomagnetic activity level, irrespective of from where it is measured. In practice, the range limits for all $K$ bands are all set by just one number, $L$, the lower limit of the $K = 9$ band because the same relative scale is used at all stations and so the thresholds for the $K$ bands 1–8 are scaled from $L$, the lower limit for the $K = 0$ band being set to zero (Menvielle & Berthelier, 1991). The derivation of the $K$ values (and from them the $a_K$ value and $aa_N$ and $aa_S$) is illustrated schematically in Figure 1.

The value of $L$ used for a station is set by its closest proximity to a nominal auroral oval. To understand this, we note that mid-latitude range indices respond most strongly to the substorm current wedge (e.g. Saba et al., 1997; Lockwood, 2013), resulting in very high correlations with auroral electrojet indices such as $AE$ and $AL$ (e.g. Adebesin, 2016). For example, the correlation coefficient between the available coincident 50 annual means of the standard auroral electrojet $AE(12)$ index and the $ap$ index (based on the $K$ values for a network of stations) is 0.98 (significant at the 99.99% level), and the correlation between the 17461 coincident daily means of $AE(12)$ and $Ap$ ($Ap$ being daily means of $ap$) is 0.84 (significant to the same level). This means that the range response of a station is greatest in the midnight Magnetic Local Time (MLT) sector (Clauer & McPherron, 1974). As well as the response being smaller away from midnight, the typical time variation waveform also varies with MLT (Caan et al., 1978). The range variation in a substorm is generally greatest in the auroral oval and decreases with decreasing latitude. This is mainly because the response of high-time-resolution geomagnetic measures (such as the $H$ component at the ground or the equivalent currents at 1-minute resolution) show a marked decrease in amplitude with increasing distance from the auroral oval (an example of the former is presented by Rostoker (1972) and a statistical survey of the latter during 116 substorms seen from 100 geomagnetic stations is presented by Gjerloev & Hoffman (2014)). This means that the range in the $H$ values in 3-hour intervals also shows a decrease with increasing distance from the auroral oval. However, we note that at lower latitudes the variation becomes rather more complex. Ritter & Lühr (2008) surveyed the effects of 4000 substorm responses statistically at 4 stations, the most poleward of which was Niemegk. They found (their Fig. 8) that the initial response to substorm expansion phase onset in 1-minute $H$ values is actually almost constant with latitude at these low and middle latitudes, but at the higher magnetic latitude stations there was a faster subsequent decay in the substorm perturbation to $H$. The resulting effect on the values of the range in $H$ during 3-hour intervals is again a tendency for them to decrease with decreasing latitude, but it appears to have a different origin from that seen at higher latitudes, closer to the auroral oval.

To account for the latitude variation of the range response, the value of $L$ used to set the $K$ band limits is set by the minimum distance between the station and a nominal auroral oval position. Because of the offset of the auroral oval towards the nightside, this minimum distance (quantified by the geocentric angle between the station and the point of closest approach of the nominal auroral oval, $\delta$) is set using a nominal oval at corrected geomagnetic latitude $\Lambda_{CG} = 69°$, which is an average oval location in the midnight sector where substorm expansions occur.

A key point is that in compiling the classic $aa$ index, the $L$ values have been assumed to remain constant over time for a given station, which means that the effects of secular changes in the geomagnetic field on $\delta$ have not been accounted for. Mayaud was aware of the potential for secular change in $\delta$ values but discounted it as small stating "note that the influence of the secular variation of the field on the distances to the auroral zone is such that the resulting variations of the lower limits for $K = 9$ are practically negligible at a scale of some tens of years" (Mayaud, 1968). Hence, in part, his view arose because saw $aa$ as being generated to cover the previous 100 years and did not foresee its continued extension to cover another 50 years. Being aware that the effect of secular change in the intrinsic field could not be ignored indefinitely, Chambodut et al. (2015) proposed new $\alpha_{15}$ indices, constructed in a way that means that the secular drift in the magnetic latitude of the observatories used is accounted for. In addition, Mursula & Martini (2007) also noted the potential effect of secular change on the $K$-values from the Sodankylä observatory.

The approach taken to generate $aa$ is that the range data were scaled into $K$-values using the band limits set by assigned $L$ values for the stations used to generate the northern and southern hemisphere indices. These values of $L$ used by ISGI to define the $K$-band scales are 500 nT for all $aa$ stations except Canberra (CNB) for where $L = 450$ nT is used, because of its greater distance from the auroral oval. These $K$ values are then converted into $a_K$ values using a standard scale called "mid-class amplitudes", K2aK (Mayaud, 1980), given by Figure 1. However, in order to achieve intercalibration of the data from different stations, the $a_K$ values from each station were multiplied by a constant correction factor for that station to give $aa_N$ and $aa_S$ for the northern and southern hemisphere, respectively. The correction factors took into account two things: a constant magnetic latitude correction and an induction effect correction. The correction factors adopted were: 1.007 for Greenwich; 0.934 for Abinger; 1.059 for Hartland; 0.967 for Melbourne; 1.033 for Toolangi; and 0.976 for Canberra (using $L = 450$ nT for Canberra). Note that this has an effect on the allowed quantization levels of the indices. Without the correction factors there would be 10 allowed levels for both $aa_N$ and $aa_S$. Averaging them together to get $aa$ would give 19 possible values. Using the scaling factors means that at any one time there are still only 19 possible quantized levels, but those levels change a little with each station change (i.e. at 1920, 1925, 1957, and 1980).

Having the two $aa$ stations roughly 10 h of local time apart means that one of the two is on the nightside at any time. This means that we cannot expect the two stations to agree at any given time. However, ideally there would be no systematic hemispheric asymmetries and, on average, the behavior of $aa_N$ and $aa_S$ should be the same. It has long been recognized that this is not the case for the classic $aa$ index. Bubenik & Fraser-Smith (1977) studied the overall distributions of $aa_N$ and $aa_S$ and found that they were different: they argued that the problem was introduced by using a quantization scheme, a potential problem discussed by Mayaud (1980). Love (2011) investigated the difference in distributions of the $K$ values on which $aa_N$ and $aa_S$ are based. This asymmetry will be investigated in Paper 2 of this series (Lockwood et al., 2018b) using a model of the time-of-year and time-of-day response functions of the stations, allied to the effects of secular change in the main field (and associated station intercalibration issues) that are the subject of the present paper.

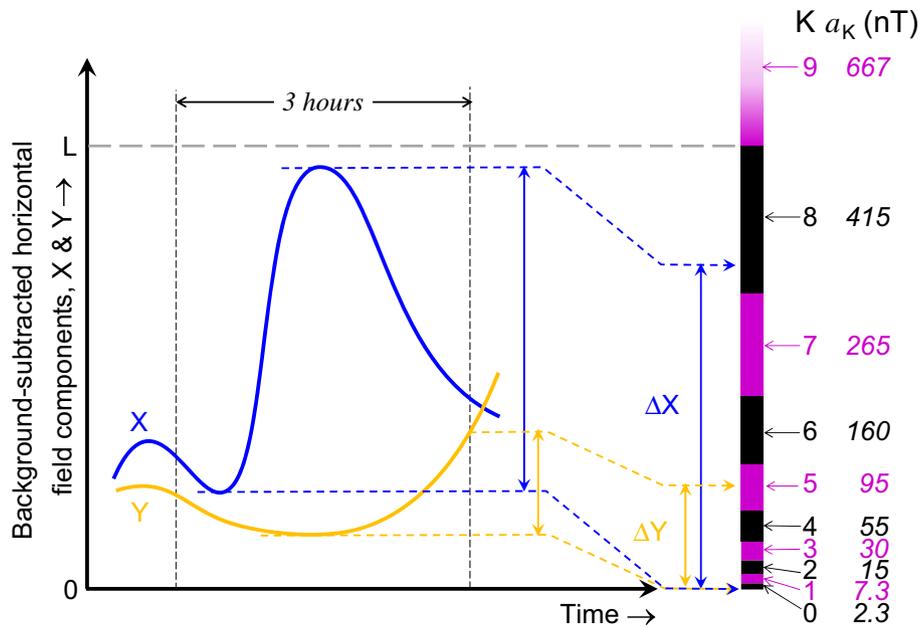

**Fig. 1.** Schematic illustration of the generation of $K$ and $a_K$ indices. Illustrative variations of the two orthogonal horizontal field components measured at one site are shown, $X$ (toward geographic north, in blue) and $Y$ (toward geographic east, in orange). These variations are after the regular diurnal variation has been subtracted from the observations. In the fixed 3-hour UT windows (00–03 UT, or 03–06 UT, and so on up to 21–24 UT), the range of variation of both components between their maximum and minimum values is taken, $\Delta X$ and $\Delta Y$. The larger value of the two is kept and scaled according to a standard, quasi-logarithmic scale (illustrated by the black and mauve bands to the right) for which all $K$-band thresholds are set for the site in question by $L$, the threshold range value for the $K = 9$ band. The value of $L$ for the site is assigned according to the minimum distance between the site and a nominal (fixed) auroral oval position. The $K$ value is then converted into the relevant quantised value of $a_K$ (in nT) using the standard "mid-class amplitudes" (K2aK) scale. In the schematic shown, $\Delta X > \Delta Y$, thus the $X$ component gives a $K$ value of 8 (whereas the $Y$ component would have given a $K$ of 5). Thus for this 3-hour interval, $a_K$ value would be 415 nT. In the case of the classic $aa$ indices, the hemispheric index ($aa_N$ or $aa_S$, for the observatory in the northern or southern hemisphere, respectively) is $f \times a_K$, where $f$ is a factor that is assumed constant for the observing site.

### 1.2 Hemispheric asymmetry in the centennial-scale change of the classic *aa* index

Figure 2a illustrates another hemispheric asymmetry in the classic $aa$ index. It shows annual means of $aa_N$ (in red) and $aa_S$ (in blue). These are the values averaged together in the generation of the official $aa$ index by L'École et Observatoire des Sciences de la Terre (EOST), a joint of the University of Strasbourg and the French National Center for Scientific Research (CNRS) institute, on behalf of the International Service of

Geomagnetic Indices (ISGI). The magnetometer data are now supplied by British Geological Survey (BGS), Edinburgh for the northern hemisphere and Geoscience Australia, Canberra for the southern hemisphere. We here refer to these $aa_N$, $aa_S$ and $aa$ data as the "classical" values, being those that are used to derive the official $aa$ index by EOST, as available from ISGI (http://isgi.unistra.fr/) and data centers around the world.

It can be seen that although $aa_N$ and $aa_S$ agree well during solar cycles 14–16 (1900–1930), $aa_N$ is progressively larger than $aa_S$ both before and after this interval. The vertical lines mark station changes (cyan for MEL to TOO; green for GRW to ABN; red for ABN to HAD; and blue for TOO to CNB). There has been much discussion about possible calibration errors between stations at these times. In particular, Svalgaard et al. (2004) pointed out that the classic $aa_N$ values showed a major change across the ABN-HAD join. These authors argued from a comparison against their "inter-hour variability" index, $IHV$, that this was responsible for an extremely large (8.1 nT) step in $aa$, such that all the upward drift in $aa$ during the 20th century was entirely erroneous. However, the early version of $IHV$ that Svalgaard et al. had employed to draw this conclusion came from just two, nearby, Northern Hemisphere stations, Cheltenham and Fredricksburg, which were intercalibrated using the available 0.75 yr of overlapping data in 1956. This calibration issue only influenced $aa_N$ and Lockwood (2003) pointed out that, as shown in Figure 2a, $aa_S$ also showed the upward rise over the 20th century, albeit of slightly smaller magnitude than that in $aa_N$ (and hence, by definition *aa*). Using more stations, Mursula et al. (2004) found there was an upward drift in *IHV* over the 20th century, but it depended on the station studied; nevertheless, they inferred that the upward drift in *aa* was probably too large. As a result, Svalgaard et al. (2003) subsequently revised their estimates of a 1957 error in *aa* down to 5.2 nT (this would mean that 64% of the drift in *aa* was erroneous). However, Mursula & Martini (2006) showed that about half of this difference was actually in the *IHV* estimates and not *aa*, being caused by the use of spot samples by Svalgaard et al., rather than hourly means, in constructing the early *IHV* data. This was corrected by Svalgaard & Cliver (2007), who revised their estimate of the *aa* error further downward to 3 nT. Other studies indicated that *aa* needed adjusting by about 2 nT at this date (Jarvis, 2004; Martini & Mursula, 2008). A concern about many of these comparisons is that they used hourly mean geomagnetic data which has a different dependence on different combinations of interplanetary parameters to range data (Lockwood, 2013). Recent tests with other range indices such as *Ap* (Lockwood et al., 2014, Matthes et al., 2016) confirm that an upward skip of about 2 nT at 1957 is present in *aa* (about one quarter of the original estimate of 8.1 nT). However, it is important to stress that this calibration arises for data which do not contain any allowance for the effects of the secular change in the geomagnetic field (in the present paper, we will show that the rise in the classic *aa* between 1902 and 1987 is indeed slightly too large, but this arises more from neglecting the change in the intrinsic geomagnetic field than from station intercalibration errors).

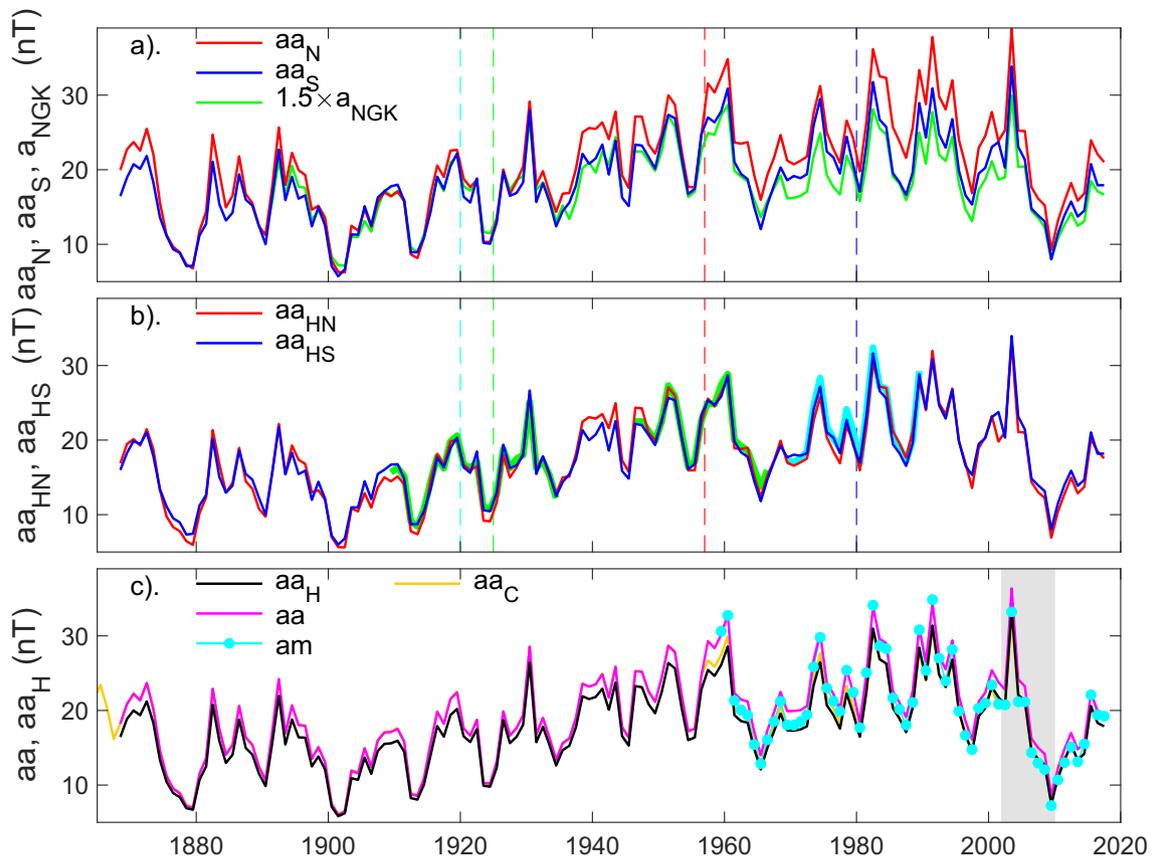

**Fig. 2.** Variations of annual means of various forms of the *aa* index. (a) The published "classic" northern and southern hemisphere indices ($aa_N$ and $aa_S$, in red and blue, respectively). Also shown (in green) is $1.5 \times a_{NGK}$, derived from the *K*-indices scaled from the Niemegk data. The vertical dashed lines mark *aa* station changes (cyan: Melbourne to Toolangi; green: Greenwich to Abinger; red: Abinger to Hartland; and blue: Toolangi to Canberra). (b) The homogenized northern and southern hemisphere indices ($aa_{HN}$ and $aa_{HS}$ in red and blue, respectively) generated in the present paper. The thick green and cyan line segments are, respectively, the $a_{NGK}$ and *am* index values used to intercalibrate segments. (c) The classic *aa* data series, $aa = (aa_N + aa_S)/2$ (in mauve) and the new homogeneous *aa* data series, $aa_H = (aa_{HN} + aa_{HS})/2$ (in black). The orange line is the corrected *aa* data series $aa_C$ generated by Lockwood et al. (2014) by re-calibration of the Abinger-to-Hartland join using the *Ap* index. (Note that before this join, *aa* and $aa_C$ are identical and the orange line is not visible as it is underneath the mauve line.) The cyan line and points show annual means of the *am* index. The gray-shaded area in (c) is the interval used to calibrate $aa_{HN}$ and $aa_{HS}$ (and hence $aa_H$) against *am*.

The argument underpinning the debate about the calibration of *aa* was that the minimum annual mean in 1901 (near 6 nT) was much lower than any seen in modern times (14 nT in 1965) and so, it was argued, erroneous. This argument as shown to be specious by the low minimum of 2009 when the annual mean *aa* fell to 8.6 nT. Furthermore, subsequent to that sunspot minimum, solar cycle 24 in *aa* has been quite similar to cycle 14 (1901–1912) and so the rise in average *aa* levels between cycles 14 and 22 has almost been matched by the fall over cycles 23 and 24. This does not necessarily mean that the classic *aa* for cycle 14 is properly calibrated, but it does mean that the frequently-used argument that it must be in error was false.

An upward 2 nT calibration skip in *aa* implies a 4 nT skip in $aa_N$ and Figure 2a shows that after 1980 $aa_N$ exceeds $aa_S$ by approximately this amount. Hence it is tempting to ascribe this difference between $aa_N$ and $aa_S$ to the one calibration skip. However, inspection of the figure reveals $aa_N$ grows relative to $aa_S$ before the ABN-HAD change in 1957. In Figure 2a, also plotted (in green) are annual mean $a_K$ values based on the *K*-index data from Niemegk (NGK, 1880–present). These have been scaled using the same mid-class amplitudes (K2aK) to give $a_{NGK}$ and then multiplied by a best-fit factor of 1.5 to bring it into line with $aa_S$. It can be seen that 1.5 $a_{NGK}$ and $aa_S$ are very similar in all years, implying that the upward drift in $aa_N$ is too large, even if it is not the ABN-HAD change that is solely responsible.

### 1.3 Studies of space climate change using the *aa* index

Feynman & Crooker (1978) reconstructed annual means of the solar wind speed, $V_{SW}$, from *aa*, using the fact that *aa*, like all range geomagnetic indices, has an approximately $V_{SW}^2$ dependence (Lockwood, 2013). However, on annual timescales, *aa* also has a dependence on the IMF field strength, *B*, which contributes considerably to the long term drift in *aa*. Lockwood et al. (1999) removed the dependence of *aa* on $V_{SW}$ using its 27-day recurrence (which varies with mean $V_{SW}$ on annual timescales) and derived the open solar flux (OSF, the total magnetic flux leaving the top of the solar corona) using "the Ulysses result" that the radial component of *B* is largely independent of heliographic latitude (Smith & Balogh, 1995; Lockwood et al., 2004; Owens et al., 2008). This variation was modelled using the OSF continuity equation by Solanki et al. (2000), who employed the sunspot number to quantify the OSF emergence rate. This modelling can be extended back to the start of regular telescopic observations in 1612. Svalgaard & Cliver (2005) noted that different geomagnetic indices have different dependencies on the IMF, *B* and the solar wind speed, $V_{SW}$, and therefore could be used in combination to derive both. This was exploited by Rouillard et al. (2007) who used *aa* in combination with indices based on hourly mean geomagnetic data to reconstruct annual means of *B*, $V_{SW}$ and OSF back to 1868. Lockwood et al. (2014) used 4 different pairings of indices, including an extended *aa* data series (with a derived 2 nT correction for a presumed $aa_N$ calibration skip in 1957) to derive *B*, $V_{SW}$ and OSF, with a full uncertainty analysis, back to 1845. Lockwood & Owens (2014) extended the modelling to divide the OSF into that in the streamer belt and in coronal holes and so computed the streamer belt width variation which matches well that deduced from historic eclipse images (Owens et al., 2017). The streamer belt width and OSF were used by Owens et al. (2017), along with 30 years' of output from a data-constrained magnetohydrodynamic model of the solar corona based on magnetograph data, to reconstruct solar wind speed $V_{SW}$ and number density $N_{SW}$ and the IMF field strength *B*, based primarily on sunspot observations. Using these empirical relations, they produced the first quantitative estimate of global solar wind variations over the last 400 years and these were employed by Lockwood et al. (2017) to compute the variation in annual mean power input into the magnetosphere and by Lockwood et al. (2018a) to estimate the variation in geomagnetic storm and substorm occurrence since before the Maunder minimum. The *aa* index data were also used by the CMIP-6 project (the 6th Coupled Model Intercomparison Project) to give a comprehensive and detailed set of solar forcing reconstructions for studies of global and regional climate and of space weather (Matthes et al., 2016). Vennerstrom et al. (2016) used the *aa* index to investigate the occurrence of great geomagnetic storms since 1868.

Hence the *aa* index has been extremely valuable in reconstructing space climate, and in taking the first steps towards a space weather climatology that covers more general conditions than do the direct satellite observations (which were almost all recorded during the Modern Grand Maximum (Lockwood et al., 2009)). In addition, the *aa* data have been hugely valuable in facilitating the exploitation of measured abundances of cosmogenic isotopes, $^{14}C$, $^{10}Be$ and $^{44}Ti$ (Usoskin, 2017). These records of past solar variability, stored in terrestrial reservoirs such as tree trunks, ice sheets and fallen meteorites, do not overlap much (or at all) with modern spacecraft data. For example, $^{14}C$ cannot be used after the first atomic bomb tests, and recent $^{10}Be$ data is less reliable as it is taken from the firn rather than the compacted snow of the ice sheet, whereas $^{44}Ti$ accumulates in meteorites over very long intervals. The extension of spacecraft data by reconstructions based on *aa* has given an overlap interval since 1868 which can be used to aid the interpretation of the cosmogenic data (Asvestari & Usoskin, 2016; Owens et al., 2016).

### 1.4 Making a homogeneous *aa* index

From Section 1.3, it is apparent that the *aa* index is very important to studies of past space climate. The issues (such as hemispheric asymmetries and calibration glitches) in the *aa* index discussed here and other limitations (such as the strong artefact diurnal variation caused by the use of just 2 stations) will not invalidate the space climate work that has been done using *aa*, although they may call for some corrections. However, the increasing use and importance of *aa* makes it timely to take a comprehensive look at these issues. In Paper 2 (Lockwood et al., 2018b) we study how the compilation of the *aa* index influences its time-of-day and time-of-year response and, as far as is possible, we make corrections for this and explain and correct the north-south asymmetries in the distributions of 3-hourly *aa* values. In the present paper, we study the difference in the long-term drift of the northern and southern *aa* indices. We show that the intercalibration glitches in *aa*, particularly that between Abinger and Hartland, were actually not just errors, but were also necessary to compensate for the drifts introduced into the data by the secular change in the intrinsic geomagnetic field. Figure 2b shows the end result of the process detailed in the present paper – a process that makes allowance for the effects of these drifts on the $aa_N$ and $aa_S$ values and then re-calibrates the joins between data from the different stations. It can be seen from Figure 2 that the resulting "homogenized" $aa_{HN}$ and $aa_{HS}$ indices obtained from this process are much more similar to each other than are the classic *aa* indices, $aa_N$ and $aa_S$.

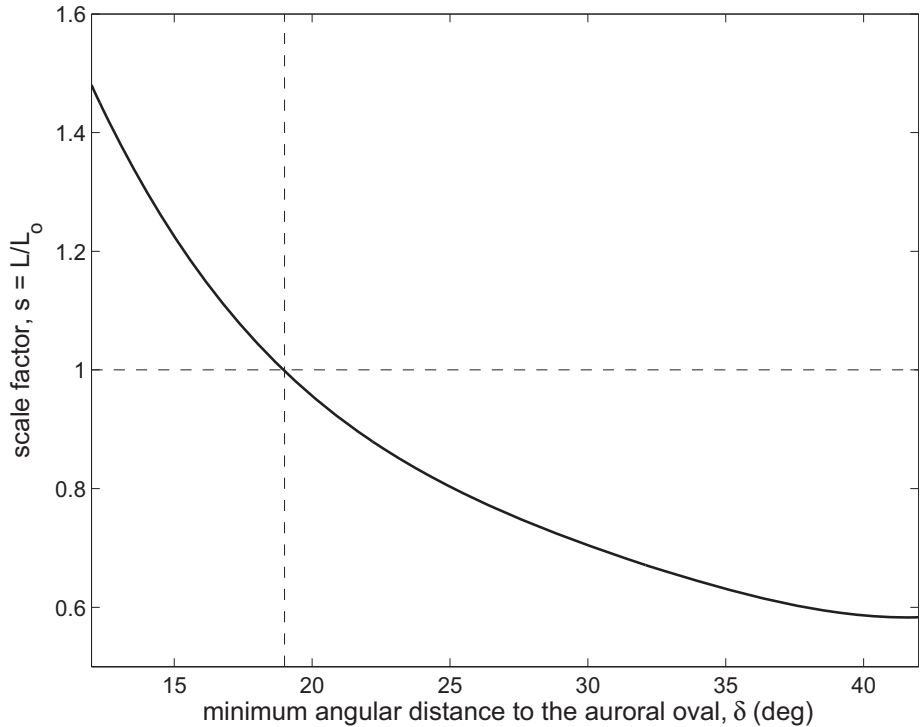

**Fig. 3.** The variation of the scale factor $s(\delta)$ derived from threshold range value $L$ that defines the $K = 9$ band, with the minimum angular separation of the station from a nominal auroral oval, $\delta$. This empirical variation is scaled from Mayaud (1968, 1972) and is the basis of the $L$ values used to scale $K$-indices from observed range for all mid-latitude stations. The scale factor $s(\delta)$ normalizes to the idealized Niemegk station for which $\delta = 19°$ and $L = Lo = 500$ nT (ideal static Mayaud values).

Note that in this paper, we do just two things. Firstly, we correct Mayaud's derivation to allow for secular drift in the main geomagnetic field – a factor which he understood but decided could be neglected. Indeed, part of the brilliance of Mayaud's formulation was to use the minimum distance to the auroral oval, which is less subject to secular change than the geomagnetic latitude of the station. This is because both the geomagnetic latitude of the station and the geographic latitude of the average auroral oval drift with the secular change in and, although the two do not change in precisely the same way, there are similarities and so part of the secular drift is cancelled out by taking the difference between the two, $\delta$. (Of course they do not cancel completely and that is why there is still a requirement to correct for the secular change in the main field). Secondly we revisit the inter-calibration of the stations which becomes necessary when the station data has been corrected for the effect of the secular field change. We take the opportunity to calibrate the revised $aa$ to modern data from the $am$ index which is derived from a global network of 24 stations. As a test of the validity of our approach we show that it makes the variations of the annual means of the northern and southern hemisphere $aa$ indices, $aa_N$ and $aa_S$, much more similar although we make no changes that were designed in advance to make them similar. The reason why this is a useful improvement to the index comes from the rationale for averaging $aa_N$ and $aa_S$ together to get an index ($aa$) that is hoped to be global in its application and implications. In deriving $aa$, Mayaud selected the sites to be as close to antipodal as possible and give a continuous data sequence: he did not do calculations that showed that although $aa_N$ and $aa_S$ are different, the sites are in somehow special such that the difference between $aa_N$ and a true global value (that would be detected from an extensive global network) is equal and opposite to that for $aa_S$ – a condition that would guarantee that on averaging one gets a valid global mean. This being the case, the only rationale for averaging $aa_N$ and $aa_S$ to get a valid representation of a global mean is that they should the same. Note that this does not alone solve the asymmetry between the distribution of the $aa_N$ and $aa_S$ values which is investigated in Paper 2 (Lockwood et al., 2018b).

## 2 The effect of secular change in the magnetic field

Figure 3 shows the variation of the scale factor, $s(\delta)$, derived from the threshold range value $L$ that defines $K = 9$, with the minimum geocentric angular separation of the station from a nominal auroral oval, $\delta$. The oval is defined to be along typical corrected geomagnetic latitude ($\Lambda_{CG}$) of the nightside aurora of 69°. This empirical variation is taken from Mayaud (1968) and is the basis of the $L$ values used to scale $K$-indices from observed range for all mid-latitude stations. The scale factor $s(\delta)$ normalizes to an idealized Niemegk station (for which $\delta = 19°$ and $L = L_o = 500$ nT, the constant reference values established by Mayaud). The curve is described by the polynomial:

$$s(\delta) = (L/L_o) = 3.8309 - 0.32401.\delta + 0.01369.\delta^2$$
$$- (2.7711.10^{-4}).\delta^3 + (2.1667.10^{-6}).\delta^4 \quad (1)$$

where $\delta$ is in degrees. Equation (1) applies over the range $11° < \delta < 40°$ which requires that the station be at mid-latitudes (the relationship not holding for either equatorial or auroral stations).

In this paper, corrected geomagnetic latitudes ($\Lambda_{CG}$), and Magnetic Local Times (MLT), are computed using the IGRF-12 model (Thébault et al., 2015) for dates after 1900. For dates before this (not covered by IGRF-12) we employ the historical gufm1 model (Jackson et al., 2000), values being scaled using linear regression of values from IGRF-12 for an overlap intercalibration interval of 1900–1920. Figure 4a shows the variations of $|\Lambda_{CG}|$ for the various stations used to generate $aa$, plus that of Niemegk (NGK, in orange). The vertical lines show the dates of transfer from one station to the next, using the same color scheme as Figure 2. It can be seen that for much of the 20th century the geomagnetic latitude of the northern and southern hemisphere stations changed in opposite directions, with the northern stations (GRW, ABN and HAD) drifting equatorward and southern (MEL and TOO) drifting poleward. This changed around 1984 when CNB began to drift equatorward, the same direction as the northern hemisphere station at that time, HAD.

These changes in the $\Lambda_{CG}$ of stations were accompanied by changes in the geographic latitude of the nominal aurora oval at $\Lambda_{CG} = 69°$. To compute $\delta$ or a given date, we use the geomagnetic field models to calculate the $\Lambda_{CG} = 69°$ contour in the elevant hemisphere in geographic coordinates and then spherical geometry to find the angular great circle distances between the station in question and points on this contour: we then iterate the geographic longitude of the point on the contour until the minimum angular distance is found, which is $\delta$. The variations of $\delta$ derived this way for each station are shown in Figure 4b. Using equation (1), this gives the variation of scale factors $s(\delta)$ in Figure 4c for each station. It can be seen that the secular change in the intrinsic field has caused a considerable drift in the threshold value for the $K = 9$ band, $L$, that should have been used. In compiling the original $aa$ index, it was assumed that $s(\delta)$ for each station remained constant (the scale factors given in Section 1.1 being $1/s(\delta)$ and assumed constant. Remember also that larger $s(\delta)$ means a higher $L$ which would give a lower $aa$ value. We could consider reanalyzing all the range data using $K$-scale band thresholds that varied according to Figure 4c: correcting the band thresholds would change many $K$-values, but would also leave many unchanged. However, there are now 150 years of $aa$ data which gives 0.87 million 3-hourly intervals to analyse from the two stations, many of which are not available as digital data. Clearly this would be a massive undertaking but it would also be a change in the construction philosophy because $aa$ values have been scaled using constant $L$ values (500 nT for all stations except Canberra for which 450 nT is used). The station correction factors applied in constructing the classic $aa$ values include an allowance for the fact that the $L$ values used are not optimum for the station in question: however, where in the classic

$aa$ these factors are constants over time, we here vary them to allow for the secular change in the intrinsic geomagnetic field. Therefore we divide classic $aa_N$ and $aa_S$ values by the $s(\delta)$ that applies for that station at that date. From the above, we stress that this type of correction is already employed in the classic $aa$ data, as it is the same principle as adopted when applying the scale factors for the station. The only difference is that here we use the IGRF-12/gufm1 model spline to apply time-dependent scale factors, $s(\delta)$, rather than the constant ones for each station used in deriving the classic $aa$.

Introducing these time-dependent scaling factors reduces the rise in $aa_N$ by 4.11%, over the interval of the Greenwich data (compared to a constant factor) – a rate of drift of 0.0721% p.a.; by 0.83% over the interval of the Abinger data (0.0258% p.a.) and by 5.37% over the interval of the Hartland data (0.0895% p.a.). On the other hand, they increase the rise in $aa_S$ by 4.77% over the interval of the Melbourne data (0.0917% p.a.); by 5.28% over the interval of the Toolangi data (0.0880% p.a.); but decrease the rise in $aa_S$ over the interval of the Canberra data by 1.84% (0.0497% p.a.). Thus allowing for the secular change in the intrinsic magnetic field reduces the disparity in the long term-drifts in $aa_N$ and $aa_S$ that can be seen in Figure 2a.

Figure 5 summarizes the differences between the computation of the classic $aa$ index and that of the new homogenized indices presented in this paper. The left-hand plots compare the variations in the minimum angular distance of the stations to the auroral oval $\delta$ and compares them to the constant values used in generating the classic $aa$ index. The right-hand plots show the corresponding scale factors, $s(\delta)$. The (constant) correction factors used in constructing $aa$ were derived account for several factors in addition to $\delta$ and their reciprocals are shown in the right-hand plots as dot-dash lines. (Reciprocals are plotted because the correction factors were multiplicative whereas we divide by the $s(\delta)$ scale factors).

The Mayaud latitude correction formulation has also been used to generate the $am$, $an$ and $as$ indices since their introduction in 1959. In generating new 15-minute indices in four local time sectors, Chambodut et al. (2015) used a different approach employing a polynomial in the stations' geomagnetic latitudes. Although the purpose of the two schemes is the same, a comparison cannot be made between them because the new Chambodut et al. (2015) indices are 15-minute range values, as opposed to the 3-hour range ($K$ index) values used by the $aa$, $am$, $as$ and $an$ indices. There are four separate indices in the Chambodut et al. (2015) set, one for each of four Magnetic Local Time (MLT) sectors whereas the Mayaud formulation is designed to account predominantly for the midnight sector by taking the minimum geomagnetic latitude offset to the auroral oval (which occurs in the midnight sector). The advantage of using geomagnetic latitude is that greater precision can be obtained (because there is no need to employ a nominal oval location) but the station calibration factor needs considerable annual updates because of the secular drift in the station's geomagnetic latitude. On the other hand, the Mayaud formulation has the advantage of being less influenced by secular change in the main field, as discussed above.

We here use Mayaud's formulation to correct for secular change via division by the $s(\delta)$ factors. However, was also taking the opportunity to re-calibrate (via linear regression) the *aa* index against the *am* index which is based on 14 stations in the northern hemisphere and 10 stations in the south. This recalibration is carried out in Section 3.1 for the Hartland and Canberra data using linear regression over 2002–2009 (inclusive), and then passed back ("daisy-chained") to earlier stations (from Hartland to Abinger and then Greenwich in Sections 3.2 and 3.3 and from Canberra to Toolangi and then to Melbourne in Section 3.4). Figure 6 demonstrates how well this approach works by (top panel) comparing the results of applying this procedure to modern $a_K$ data from a range of stations at different geographic latitudes, $\lambda_G$: (mauve) Sodankylä, SOD, $\lambda_G = 67.367°N$; (brown) Eskdalemuir, ESK, $\lambda_G = 55.314°N$; (orange) Niemegk, NGK, $\lambda_G = 52.072°N$; (red) Hartland, HAD, $\lambda_G = 50.995°N$; (blue) Canberra, CNB, $\lambda_G = 35.315°S$; and (green) a spline of Gangara, GNA, $\lambda_G = 31.780°S$ and nearby Gingin, GNG, $\lambda_G = 31.356°S$, Gingin is the replacement for Gangara after January 2013 and the spline was made using the overlap data between August 2010 and January 2013: this station pair is chosen as they are in the same southern hemisphere longitude sector as Melbourne but are at lower geomagnetic latitude (see below). The black line shows the *am* index data, the linear regression against which over the calibration interval (2002–2009 inclusive) gives the slope *m* and an intercept *i* for each station. The data are means over 27-day Bartels solar rotation intervals and cover 1995 to the present day for reasons discussed in later in this section. It can be seen that the level of agreement between the station data processed this way and the *am* calibration data is very close for all stations. The scalefactors $s(\delta)$ used in Figure 6 vary with time and location between a minimum of 0.896 (for Gangara/ Gingin) and maximum of 2.298 (for Sodankylä). The range covered by the *aa* stations is 0.940 (for Melbourne in 1875) and 1.102 (for Greenwich in 1868) – hence our test set of stations covers all of the range of $\delta$ for the *aa* stations, plus a considerable amount more. The bottom panel of Figure 6 shows the root-mean-square (rms) deviation of the individual station values from the *am* index, $\varepsilon_{rms}$. For most Bartels' rotations this is around 5%, but in the low solar minimum of 2008/2009 rises to consistently exceed 15% and in one 27-day interval reaches almost 50%. This is partly because these are percentage errors and the values of *am* are low, but also because by averaging 24 stations, *am* has much greater sensitivity at low values than $a_K$ values from a single station. For these 27-day intervals the mean $\varepsilon_{rms}$ is 9.2% and this is reduced to 3.1% in annual mean data. Hence the procedure we deploy makes modern stations give, to a very good degree of accuracy, highly consistent corrected $a_K$ values, even though they cover a much wider range of $\delta$, and hence correction factors $s(\delta)$, than are covered by the *aa* stations since the start of the *aa* data in 1868. We estimate that for the range of $s(\delta)$ involved in the historic *aa* data, the latitudinal correction procedure for annual means is accurate to better than 1% on average.

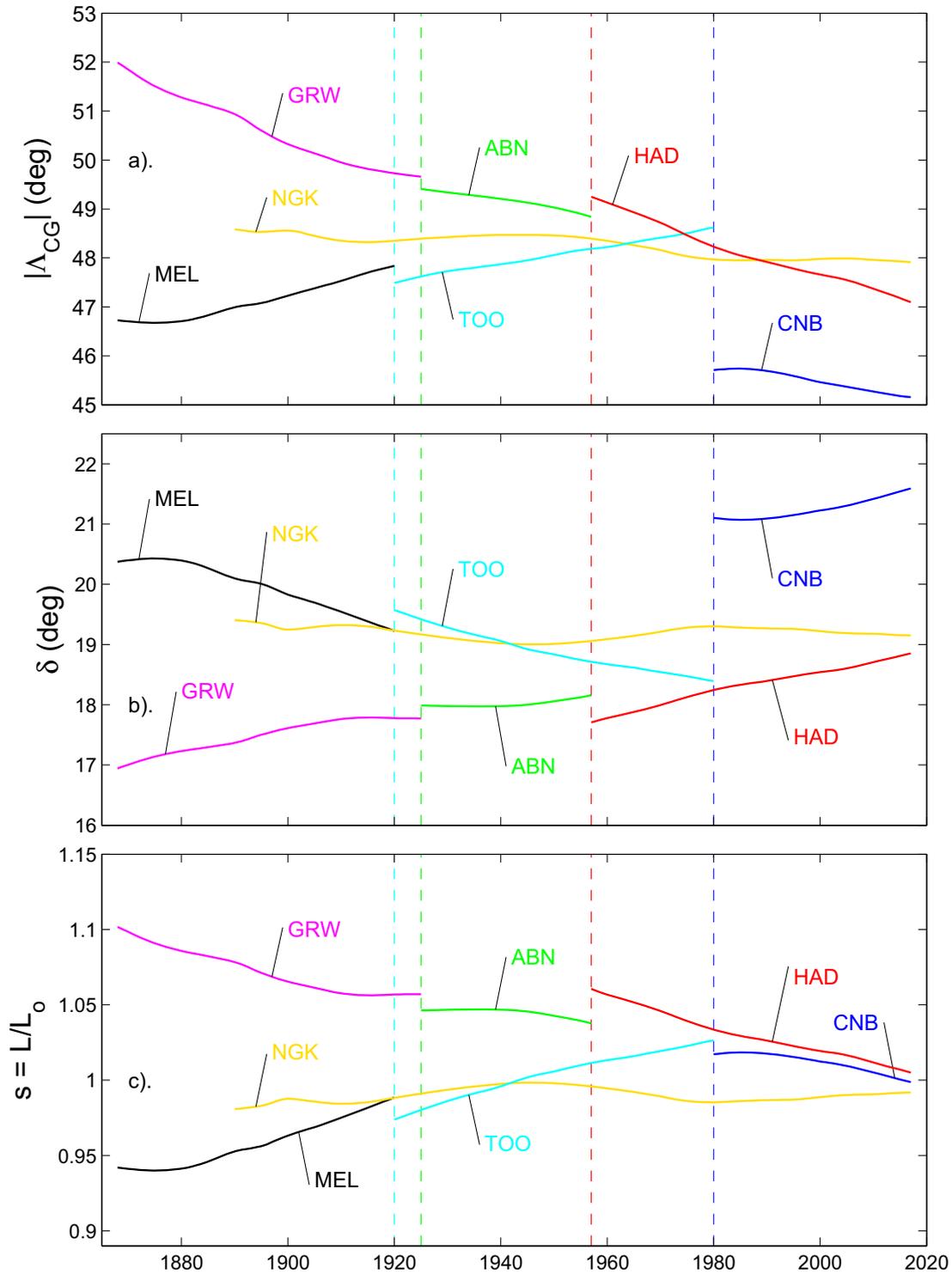

**Fig. 4.** Analysis of the effect of secular change in the geomagnetic field on the *aa* magnetometer stations using a spline of the IGRF-12 and the gufm1 geomagnetic field models (for after and before 1900, respectively). (a) The modulus of the corrected geomagnetic latitude, $|\Lambda_{CG}|$ of the stations; (b) the angular separation of the closest approach to the station of a nominal nightside auroral oval (at $|\Lambda_{CG}| = 69°$), $\delta$; and (c) the scale factor $s(\delta) = L/L_o$ where $L$ is given as a function of $\delta$ by Figure 3 and $L_o = 500$ nT, the reference value for the Niemegk station (for which $\delta$ is taken to be 19°) except for Canberra which, because of its more equatorward location, is scaled using $L_o = 450$ nT. The northern hemisphere stations are Greenwich (code GRW, in mauve), Abinger (ABN, in green) and Hartland (HAD, in red). The southern hemisphere stations are Melbourne (MEL, in black), Toolangi (TOO, in cyan) and Canberra (CNB, in blue). Also shown is Niemegk (NGK, in orange: data available since 1890). Vertical dashed lines mark *aa* station changes.

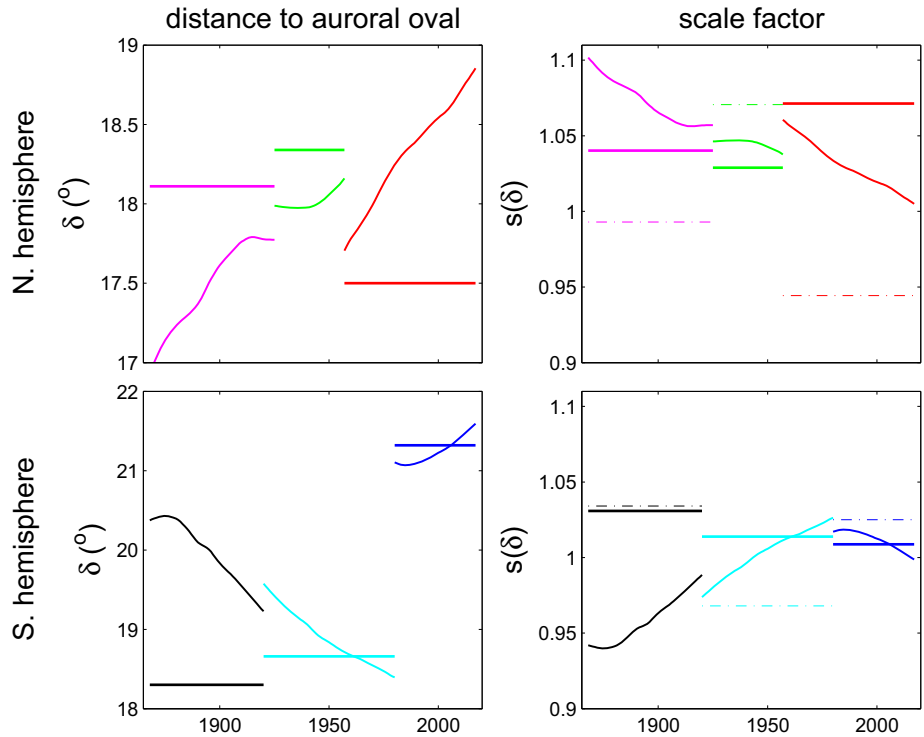

**Fig. 5.** Variations of (left) the minimum angular distance to the auroral oval, $\delta$, and (right) the scalefactors, $s(\delta)$, for the *aa* stations. The colours used are as in Figure 4 (namely mauve for Greenwich, green for Abinger, red for Hartland, black for Melbourne, cyan for Toolangi and blue for Canberra). The thin lines are the variations shown in Figure 4 and the thick lines are constant values used in generating the classic *aa*. The dot-dash lines in the right-hand panels show the reciprocals of the standard multiplicative correction factors and the thick lines the factors corresponding to the constant $\delta$ values in the left-hand panels.

As discussed in the introduction, a major application of the *aa* index is in reconstructing the near-Earth interplanetary conditions of the past and so it is useful to evaluate if the errors shown in Figure 6 are significant in this context. The data in

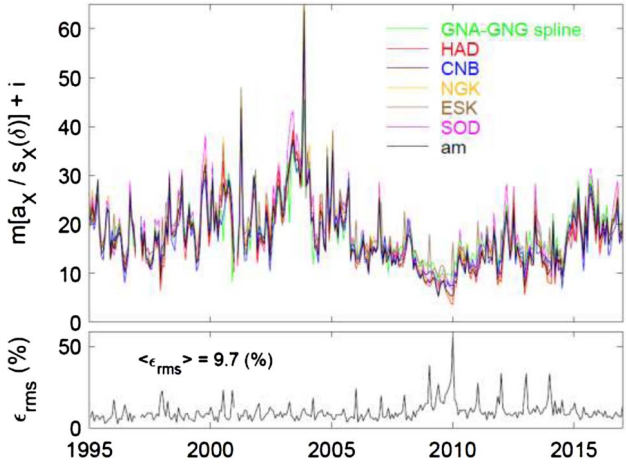

**Fig. 6.** Top: Scaled variations of modern $a_K$ values from various stations using the station location correction procedure used in this paper. For all stations, the observed $a_K$ values have been corrected for any secular magnetic field change by dividing by the $s(\delta)$ factor and then scaled to the $am$ index using the linear regression coefficients $m$ and $i$ obtained from the calibration interval (2002–2009, inclusive). The plot shows 27-day Bartels rotation means for data from: (mauve) Sodankylä, SOD; (brown) Eskdalemuir, ESK; (orange) Niemegk, NGK; (red) Hartland, HAD; (blue) Canberra, CNB; and (green) a spline of Gangara, GNA and nearby Gingin, GNG (see text for details). The black line is the $am$ index. Bottom: the rms. fit residual of the re-scaled station $a_K$ indices compared with the $am$ index, $\varepsilon_{rms}$, for the 27-day means. The average of $\varepsilon_{rms}$ for the whole interval shown (1995–2017), is $\langle \varepsilon_{rms} \rangle = 9.7\%$

Figure 6 are restricted after 1995 because this allows to make comparisons with near-continuous data from near-Earth interplanetary space. Lockwood et al. (2018c) have shown that gaps in the interplanetary data series render most "coupling functions" (combinations of near-Earth interplanetary parameters used to explain or predict geomagnetic disturbance) highly inaccurate if they are derived using data from before 1995. By introducing synthetic data gaps into near-continuous data, these authors show that in many cases differences between derived coupling functions can arise because one is fitting to the noise introduced by the presence of many and long data gaps. After 1995 the WIND, ACE and DISCOVR satellites give much more continuous measurements with fewer and much shorter data gaps. Because of the danger of such "overfitting", Lockwood et al. (2018c) recommend the power input into the magnetosphere, $P_\alpha$, as the best coupling function. This is because $P_\alpha$ uses the theoretical basis by Vasyliunas et al. (1982) to reduce the number of free fit variable to just one, the coupling exponent $\alpha$, and yet achieves almost as high correlations with range geomagnetic indices as coupling functions that have separate exponents for different solar wind variables which, if they do achieve a slightly higher correlation, tend to do so by overfitting and with reduced significance because of the increased number of free fit parameters. The equation for $P_\alpha$ shows a dependence on $B^{2\alpha}V_{SW}^{(7/3-\alpha)}(m_{sw}N_{sw})^{(2/3-\alpha)}$ (where $B$ is the interplanetary magnetic field $V_{SW}$ is the solar wind speed and $(m_{sw}N_{sw})$ is the mass density on the solar wind) and so accounts for all three near-Earth interplanetary parameters with one free fit parameter, the coupling exponent, $\alpha$. This is much preferable to forms such as $B^a V_{SW}^b(m_{sw}N_{sw})^c$ which have three free fit parameters and so are much more prone to "overfitting".

In evaluating $P_\alpha$, great care is here taken in handling data gaps because the often-used assumption that they have no effect on correlation studies can be a serious source of error. As pointed out by Lockwood et al. (2018c), the much-used Omni2 interplanetary dataset gives an hourly mean value even if there is just one sample available within the hour. This is adequate for parameters such as $V_{SW}$ that have high persistence (i.e. long autocorrelation timescales) but inadequate for parameters such as the IMF orientation factor that has and extremely short autocorrelation timescale. Another complication is that, although coupling functions made by averaging interplanetary parameters and then combining them are valid and valuable, they are not as accurate as ones combined at high time resolution and then averaged. Hence we here start from 1-minute Omni data (for after 1995 when data gaps are much fewer and shorter). Hourly means of a parameter are then constructed only when there are sufficient 1-minute samples of that parameter to reduce the uncertainty in the hourly mean to 5%. The required number of samples for each parameter was obtained from the Monte-Carlo sampling tests carried out by Lockwood et al. (2018c). From these data, hourly means of $P_\alpha$ are constructed (for a range of $\alpha$ values between 0 and 1.25 in steps of 0.01). Note that a data gap in the $P_\alpha$ sequence is formed if any of the required parameters is unavailable. These hourly $P_\alpha$ samples are then made into 3-hourly means (matching the 8 time-of-day intervals of the geomagnetic range indices) only when all three of the required hourly means of $P_\alpha$ are available. Lastly, as used by Finch & Lockwood (2007), each geomagnetic index data series is masked out at times of the data gaps in the 3-hourly $P_\alpha$ samples (and the $P_\alpha$ data correspondingly masked out at the times of any gaps in the geomagnetic data it is being compared to) so that when averages over a longer interval are taken (we here use both 27-day Bartels solar rotation intervals and 1-year intervals) only valid coincident data are included in the averages of both data sets to be correlated. We find this rather laborious procedure improves the correlations and removes many of the apparent differences between the responses of different geomagnetic observatories.

Figure 7 shows the resulting correlograms for the Bartels rotation (27-day) means for the stations also used in Figure 6. The correlation coefficient is shown as a function of the coupling exponent, $\alpha$. The peak correlations for these 27-day means are of order 0.93 and rise to over 0.98 for annual means. Using the three separate exponents $a$, $b$ and $c$ (discussed above) causes only very small increases in the peak correlation that are not statistically significant when one allows for the additional number of degrees of freedom. The optimum exponent for $am$ for the 27-day means is $\alpha = 0.45 \pm 0.07$ (see Lockwood et al. (2018c) for description of the two error estimation techniques that are used to generate these 1-$\sigma$ uncertainties) giving a peak correlation of 0.93. For annual means the peak correlation for $am$ is 0.99 at $\alpha = 0.44 \pm 0.02$ (Lockwood et al., 2018c). The optimum values for all but two of the $a_K$ stations tested fall in, or close to, this range (shown by the coloured dots and vertical dashed lines). The optimum $\alpha$ for

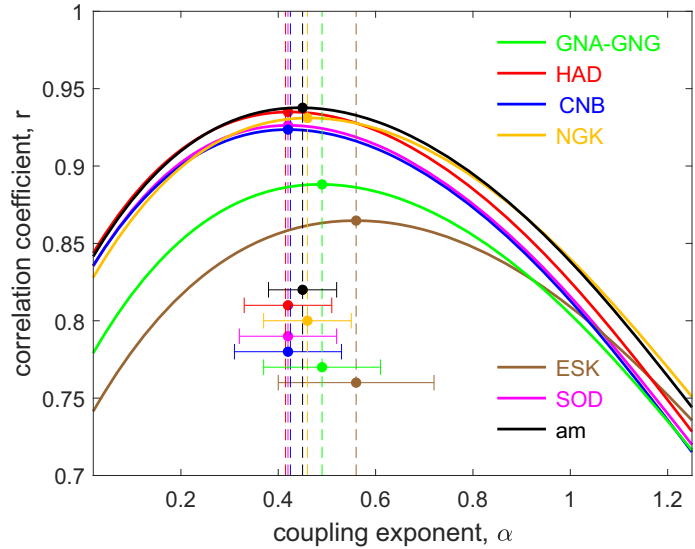

**Fig. 7.** Correlograms showing the correlation between 27-day Bartels solar rotation means of power input into the magnetosphere, $P\alpha$, with the corrected $a_K$ indices, $a_K/s(\delta)$, as a function of the coupling exponent, $\alpha$. The colours are for the same data as used in Figure 6: (mauve) Sodankylä, SOD; (brown) Eskdalemuir, ESK; (orange) Niemegk, NGK; (red) Hartland, HAD; (blue) Canberra, CNB; and (green) a spline of Gangara, GNA and nearby Gingin, GNG (see text for details). The black line is the *am* index. The coloured dots and vertical dashed lines show the optimum $\alpha$ that gives the peak correlation. The horizontal bars show the uncertainty in the optimum $\alpha$ which is the larger of the two 1-$\sigma$ uncertainties computed using the two procedures described by Lockwood et al. (2018c).

Sodankylä ($0.42 \pm 0.10$, in mauve), Niemegk ($0.46 \pm 0.09$, in orange), Hartland ($0.42 \pm 0.09$, in red), Canberra ($0.42 \pm 0.11$, in blue), for Gangara/Gingin ($0.49 \pm 0.12$, in green) and Eskdalemuir ($0.56 \pm 0.16$, in brown) all agree with that for *am* to within the estimated uncertainties and all show considerable overlap in estimated uncertainty range with that for *am*. Note that the peak correlation coefficient is also considerably lower for ESK and we find, in general, that increased geomagnetic station noise, and in particular lower instrument sensitivity, increases the optimum $\alpha$ (and its uncertainty range) as well as lowering the peak correlation. We find no consistent variation with magnetic latitude nor with the minimum distance to the auroral oval, $\delta$ and effectively the same coupling exponent applies at Sodankylä (considerably closer to the auroral oval than any of the *aa* stations at any date) as at Gangara/Gingin (further away from the auroral oval than any *aa* stations at any date). Hence this test shows that the changing magnetic latitudes of the *aa* stations is not introducing long-term changes into the response of the index to interplanetary conditions.

## 3 Recalibrating the stations

The drift in the scaling factors will have influenced the intercalibration of the stations. Consider the Abinger-Hartland join in 1957, which has been the cause of much debate, as discussed in Section 1.2. By end of the interval of the Abinger data, the use of a constant scale factor means that the classic *aa* was giving $aa_N$ values that were too high by $1.44/2 = 0.72\%$, compared to the mean value for the Abinger interval. On the other hand, for the start of the Hartland data, classic $aa_N$ values were too low compared to the average for the Hartland interval by 4.41%. Given that the average $aa_N$ value was 24.6 nT for 1956 and 31.6 nT for 1957, this makes a difference of 1.6 nT which is approximately half that required to explain the apparent calibration skip between the Abinger and Hartland data. This throws a new light on the calibration "glitch" at the ABN-HAD join which can be regarded as being as much a necessary correction to allow for the effect of the drift in the intrinsic magnetic field as a calibration error.

If we knew the precise dates for which the classic *aa* index (constant) scalefactors applied, we could generalize them using the $s(\delta)$ factors and so employ Mayaud's original station inter-calibrations. However, these dates are not clear and so the corrected indices $aa_N/s(\delta)$ and $aa_S/s(\delta)$ need new intercalibrations, which is done in this section using independent data. We take the opportunity to make calibrations that can also allow for other potential factors, such as any change in the subtraction of the regular diurnal variation associated with the change from manual to automated scaling. For both the two northern hemisphere station changes we use data from the Niemegk (NGK) station in Germany, $K$ indices from where are available from 1890. Figure 4c shows that the $s(\delta)$ factor is relatively constant for NGK (orange line) but there are nevertheless some small changes (the range of variation in $s(\delta)$ for NGK in Figure 4c is 1.8%). Hence we use $a_{NGK}/s(\delta)$, where $a_{NGK}$ is scaled from the NGK $K$ values using the standard mid-class amplitudes scale (K2aK). For the southern hemisphere we have no independent $K$-index record that is as long, nor as stable, as that from NGK. For the Toolangi-Canberra join, we use the *am* index (compiled for a network of stations in both hemispheres, Mayaud, 1980; Chambodut et al., 2013), but find we get almost identical results if we use the southern hemisphere component of *am*, *as*, or its northern hemisphere component, *an*, or even $a_{NGK}/s(\delta)$. For the Melbourne-Toolangi join we have no other data of the duration and quality of Niemegk and so we use use $a_{NGK}/s(\delta)$.

The procedure used is to take 11 years' data from each side of the join (roughly one solar cycle). For both the "before" and "after" interval we compare the *aa* station data with the calibration station data. We employ daily means, thereby averaging out the diurnal variations. As discussed in the next paragraph, we carry out the calibration separately for eight independent equal-length time-of-year ($F$) ranges in which we regress the corrected *aa* station data against the corrected calibration set (for the 11 years before and after the join, respectively). This means that each regression is carried out on approximately 500 pairs of daily mean values ($11 \times 365/8$). All regressions were tested to ensure problems did not arise because of lack of homoscedacity, outliers, non-linearity, interdependence and using a Q-Q test to ensure the distribution of residuals was Gaussian (thereby ensuring that none of the assumptions of ordinary least squares regression, OLS, are violated). The scatter plot was also checked in the 11 annual-mean data points because the main application of the regressions in this paper is to annual means. The "before" and "after" regressions were then compared, as discussed below.

There are a number of reasons to be concerned about seasonal variation in magnetometer calibrations. These may be instrumental, for example early instruments were particularly temperature and humidity sensitive. In addition, ionospheric Earth currents can depend on the height of the water table (although their effect is predominantly in the vertical rather than the horizontal components). In the case of Hartland, its coastal location makes ocean currents, and their seasonal variation, a potential factor. All these may differ at different sites. The conductivities of the ionosphere, and their spatial distribution above the station, and between the station and the auroral oval, will have a strong seasonal component and again this factor may not be exactly the same at different sites. Possibly the largest concern is the quiet-time regular variation, $S_R$, that must be subtracted from the data before the range is evaluated and this correction may vary with season as the $S_R$ pattern moves in location over the year (Mursula et al., 2009). We note that Matthes et al. (2016) used the *Ap* index, derived from a wider network of mid-latitude magnetometers, to re-calibrate the Abinger-Hartland join in the $aa_N$ data and found that the calibration required varied with time-of-year. For this reason, the calibrations are carried out separately in the 8 independent time-of-year ($F$) bins: the number of $F$ bins was chosen as a compromise between resolution of any annual variation and maintaining a high number of samples in each regression. Although, there was general agreement between the results from the different $F$ bins, there were also consistent differences at some times of year. Note that this procedure allows us to re-calibrate not only instrumental effects but also any changes in the background subtraction and scaling practices used to derive the $K$-indices. Scaling has changed from manual to automated and although the latter are repeatable and testable, the former are not; however, it helps increase homogeny that most of the classic *aa* data up to 1968 was scaled by Mayaud himself. Lastly, we note that Bartels recognized the need to allow for changes during the year in the intercalibration of stations because the conversion factors that he derived (and are still used to this day to derive the *Kp* index) not only depend on the station location, the Universal Time, and the activity level, but also depend on the time of year. Bartels employed 4 intervals in the year with three calibration categories (summer, winter and equinox).

By virtue of its more extensive network of stations in both hemispheres, and its use of area-weighted groupings, the *am* index is, by far, the best standard available to us for a global range index. Starting in 1959, it is coincident in time with all the Canberra data and almost all of the Hartland data. It therefore makes good sense to scale both the $aa_S/s(\delta)$ and $aa_S/s(\delta)$ data to recent *am* data, and then "daisy-chain" the calibration back to the prior two stations. As noted in the case of the sunspot number data composite (Lockwood et al., 2016), there are always concerns about accumulating errors in daisy chaining; however, we note that the calibration is here passed across only two joins in each hemisphere and the correlations with independent data used to calibrate the joins are exceptionally high. Furthermore, we have an additional check (of a kind not available to use when making the many joins needed for the sunspot number composite), namely that we have independent data from other stations (and equivalent data in the *IHV* index) that continues through much of the sequence and across all four joins. Strictly-speaking, the Niemegk data are also a composite, the data series coming from three nearby sites that are within 40 km of each other: Potsdam (1880–1907), Seddin (1908–1930), and Niemegk (1931–present). The site changes were made to eliminate the influence of local electrical noise. Of these site changes, only that in 1930 falls within the 11-year calibration periods (either side of an *aa* station change) that are deployed here, being 5 years after the Greenwich-Abinger join and 10 years after the Melbourne-Toolangi join. We note there are probably improvements that could be made to the Potsdam/Seddin/Niemegk $a_{NGK}$ composite, particularly using data from relatively nearby observatories, such as Swider (SWI), Rude Skov (RSV), Lovö (LOV) and Wingst (WNG) (e.g. Kobylinski and Wysokinski, 2006). Using local stations is preferable because the more distant they are, the larger the difference in the change in their $s(\delta)$ factors and hence the more they depend on the main field model used. Some calibration jumps in $a_{NGK}$ have been discussed around 1932 and 1996: the latter is not in an interval used for calibration in this paper, but 1932 does fall within the 22-year spline interval used to calibrate the Greenwich-Abinger join in 1925.

To test the suitability of the Niemegk $a_K$ index data for use as a calibration spline, we search for long-term drifts relative to independent data. Given that fluctuations within the 11-year "before" and "after" intervals will be accommodated by the relevant regression with the *aa* station data, our only concern is that the mean over the before interval is consistent with that over the after interval. One station that provides $K$-indices that cover all the *aa* calibration intervals is Sodankylä (SOD) from where $K$-indices data is available since 1914 and the SOD data have been used to test and re-calibrate *aa* in the past (Clilverd et al., 2005). The correlation between daily means of $a_{NGK}$ and $a_{SOD}$ exceeds 0.59 for the calibration intervals and the corresponding correlation of annual means always exceeds 0.97. However, this is not an ideal site (geographic coordinates 67.367°N, 26.633 E) in that it is closer to the auroral oval than the mid-latitude stations that we are calibrating: its $\delta$ falls from 6.11 in 1914 to 4.69 in 2017 and these $\delta$ values are below the range over which Mayaud recommends the use of the polynomial given in Equation (1). Figure 3 highlights why this a concern, as it shows that the effects of secular changes in the geomagnetic field on the required scaling factor are increasingly greater at smaller $\delta$. Equation (1) predicts that $s(\delta)$ for Sodankylä (SOD) will have risen from 2.302 to 2.586 over the interval 1914–2017, which would make the corrected SOD data more sensitive to the secular change correction than the data from lower-latitude stations. However, at this point we must remember that in applying Equation (1) to the SOD data we are using it outside the latitude range which Mayaud intended it to be used and also outside the latitude range of data that Mayaud used to derive it. However, Figure 6 shows that using Equation (1) with SOC data over two solar cycles has not introduced a serious error into the $a_{SOD}/s(\delta)$ and so it does supply a valuable additional test of the NGK intercalibration data (which also covers 2 solar cycles).

Nevertheless, because of these concerns over the $a_{SOD}/s(\delta)$ data, we have also used data from other stations, in particular the $K$-indices from Lerwick (LER) and Eskdalemuir (ESK) for the 22 years around the Abinger-Hartland join. We find it is important to correct the $K$-indices from these stations to allow for effect of changing $\delta$ because otherwise one finds false drifts relative to Niemegk, where the change in $\delta$ has been much smaller (see Fig. 4). The procedure employed here is to linearly regress $\langle a_{NGK}/s(\delta)\rangle_{\tau=1yr}$ and $\langle a_{XXX}/s(\delta)\rangle_{\tau=1yr}$, where XXX is a generic IAGA code of the station used (giving regression slope $\alpha$ and intercept $\beta$) then compare the ratio

$$M = \langle a_{NGK}/s(\delta)\rangle_{\tau=11yr} / (\alpha\langle a_{XXX}/s(\delta)\rangle_{\tau=11yr} + \beta) \quad (2)$$

for the 11-year intervals before and after ($M_B$ and $M_A$, respectively). The ideal result would be $M_A/M_B = 1$, which would mean that any change across the join in $a_{NGK}/s(\delta)$ and $a_{XXX}/s(\delta)$ was the same. Because it is highly unlikely that Niemegk and station XXX share exactly the same error at precisely the time of the join, this would give great confidence in the intercalibration.

The steps taken to generate the "homogeneous" aa indices, $aa_{HN}$, $aa_{HS}$ and $aa_H$, are given sequentially in the following subsections. It should be noted that we are using daisy chaining of calibrations which partially avoided in the classic aa index only because it was assumed that the station scale factors were constant, an assumption that we here show causes its own problems. Even then, the use of the station scale factors was, in effect, a form of daisy chaining.

### 3.1. Scaling of the Hartland and Canberra data to the am index

The first step is to remove the constant scale factors used in the compilation of the classic aa index to recover the 3-hourly $a_K$ indices, i.e. for Greenwich we compute $a_{GRW} = [aa_N]_{GRW}/1.007$, and similarly we use $a_{ABN} = [aa_N]_{ABN}/0.934$, $a_{HAD} = [aa_N]_{HAD}/1.059$, $a_{MEL} = [aa_S]_{MEL}/0.967$, $a_{TOO} = [aa_S]_{TOO}/1.033$, and $a_{CNB} = [aa_S]_{CNB}/1.084$. Given that the major application of the aa index is to map modern conditions back in time, it makes sense to scale a new corrected version to modern data. Hence we start the process of generating a new, "homogeneous" aa data series by scaling modern $a_K/s(\delta)$ data (i.e. the $a_K$ values corrected for the secular change in the geomagnetic field) against a modern standard. We use the am index as it is by far the best range-based index in terms of reducing the false variations introduced by limited station coverage and being homogeneous over time in the distribution stations it has taken data from. However, it contains no allowance for the effects of long-term change in the geomagnetic field and therefore we carry out scaling of $a_{HAD}/s(\delta)$ and $a_{CNB}/s(\delta)$ data (from Hartland and Canberra, respectively) against am for a limited period only. We employ daily means ($Am$, $A_{CNB}$ and $A_{HAD}$) to average out the strong diurnal variation in the $a_K$ indices caused by the use of just one station and the (much smaller) residual diurnal variation in am caused by the slightly inhomogeneous longitudinal coverage (particularly in the southern hemisphere) of the am stations. We use an interval of 7 years because we find that it is the optimum number to minimise estimated uncertainties: we employ 2002–2009 (inclusive) because that interval contains the largest annual mean aa index in the full 150-year record (in 2003) and also the lowest in modern times (in 2009), which is only slightly larger than the minimum in the whole record. Hence this interval covers almost the full range of classic aa values. The correlation of the daily means in this interval (23376 in number) are exceptionally high being 0.978 for $Am$ and $A_{HAD}/s(\delta)$ and 0.969 for $Am$ and $A_{CNB}/s(\delta)$. Linear regressions (ordinary least squares) between these pairs of data series pass all tests listed above and yield the scaling factors given in Table 1. In all regressions between data series we use both the slope (i.e. a gain term, $s_c$) and the intercept (an offset term, $c_c$) because, in addition to differences in instrument sensitivity, noise levels and background subtraction means that there may, in general, also be zero-level differences. Hence we scale $a_{HAD}/s(\delta)$ from Hartland using:

$$[aa_{HN}]_{HAD} = 0.9566 \cdot aa_{HAD}/s(\delta) - 1.3448$$

(for 1957–present) (3)

and we scale $a_{CNB}/s(\delta)$ from Canberra using:

$$[aa_{HS}]_{CNB} = 0.9507 \cdot a_{CNB}/s(\delta) + 0.4660$$

(for 1980–present) (4)

Over the interval 1980–present, this gives a distribution of 3-hourly ($[aa_{HN}]_{HAD} - [aa_{HS}]_{CNB}$) values with a mode value of zero, which means there is no systematic difference between the re-scaled indices from the two sites.

### 3.2. Inter-calibration of the Hartland and Abinger

Figure 8 details the method by which the Abinger data is calibrated to provide a backwards extension of the Hartland data which is as seamless as possible. As discussed above, the calibration was separated into 8 independent, equal-duration bins of the fraction of the year, $F$. Bin 1 is for $0 \leq F < 0.125$; bin 2 is $0.125 \leq F < 0.25$; and so on, up to bin 8 for $0.875 \leq F \leq 1$. The left hand column of Figure 8 shows scatter plots between the $a_{ABN}/s(\delta)$ values (i.e. the classic aa values from Abinger after removal of the original scalefactor correction and allowance for the effect of the changing intrinsic field) against the $a_{NGK}/s(\delta)$ values (the similarly-corrected values from the Niemegk $K$ indices) for the 11-year period before the join and the middle column gives the scatter plots of the corrected and re-scaled aa index values from Hartland, $[aa_{HN}]_{HAD}$, as given by Equation (2), for the 11-year period after the join, again against the simultaneous $a_{NGK}/s(\delta)$ values. In each case, the grey dots are the scatter plot for daily values and black dots are the annual means (for the range of $F$ in question). The correlation coefficients for the daily values are given in Table 1 (we do not give the corresponding correlations for annual means as they all between 0.99 and 0.999 but of lower statistical significance, coming from just 11 samples). The red lines are linear least-squares regression fits to the daily values and all tests show that this is appropriate in all cases. The third column plots the best linear fit of $a_{NGK}/s(\delta)$ in the interval after the join ("fit 2") as a function of the best linear fit of $a_{NGK}/s(\delta)$ in the interval before the join ("fit 1"). The dashed line is the diagonal and would apply if the relationship of the data before the join to $a_{NGK}/s(\delta)$ were identical to that after the join. The red lines in the right-hand column have slope $s_c$ and intercept $c_c$. Assuming that there is no discontinuity in $a_{NGK}/s(\delta)$ coincidentally at the time of the join (which means that the relationship between the calibration data and the real $aa$ index before the join is the same as that after the join) we can calibrate the Abinger data (corrected for secular drift) with that from Hartland (rescaled to $am$, as discussed in the previous section) for a given $F$ using:

$$[aa_{HN}]_{ABN}(F) = s_c(F) \cdot a_{ABN}(F)/s(\delta) + c_c(F) \qquad (5)$$

**Table 1.** The correlation coefficients ($r_b$ and $r_a$ for daily means in 11 years before and after the joins, respectively) and the slope $s_c$ and intercept $c_c$ for recalibrating stations for the 8 time-of-year ($F$) bins employed.

| Correction | Date | Fraction of year, $F$ | $F$ bin | Correlations ($\tau = 1$ day) Before, $r_b$ | After, $r_a$ | Slope, $s_c$ | Intercept, $c_c$ (nT) |
|---|---|---|---|---|---|---|---|
| Northern hemisphere | | | | | | | |
| Scale $a_{HAD}$ to $am$ | 2002–2009 | $0 \le F < 1$ | All | 0.978 | | 0.9566 | −1.3448 |
| Scale $a_{ABN}/s$ to $[aa_{HN}]_{HAD}$ | 1957 | $0 \le F < 0.125$ | 1 | 0.977 | 0.981 | 0.8629 | 0.3828 |
| | | $0.125 \le F < 0.25$ | 2 | 0.973 | 0.980 | 0.8381 | 0.9176 |
| | | $0.25 \le F < 0.375$ | 3 | 0.980 | 0.982 | 1.0112 | −1.8577 |
| | | $0.375 \le F < 0.5$ | 4 | 0.961 | 0.968 | 0.8073 | 0.7078 |
| | | $0.5 \le F < 0.625$ | 5 | 0.966 | 0.987 | 0.8274 | 0.5914 |
| | | $0.625 \le F < 0.75$ | 6 | 0.974 | 0.980 | 0.8744 | −0.0868 |
| | | $0.75 \le F < 0.875$ | 7 | 0.965 | 0.987 | 0.8820 | −0.2354 |
| | | $0.875 \le F < 1$ | 8 | 0.961 | 0.962 | 0.9315 | −1.1993 |
| Scale $a_{GRW}/s$ to $[aa_{HN}]_{ABN}$ | 1925 | $0 \le F < 0.125$ | 1 | 0.958 | 0.968 | 0.8247 | 1.0065 |
| | | $0.125 \le F < 0.25$ | 2 | 0.968 | 0.967 | 0.9650 | −0.2352 |
| | | $0.25 \le F < 0.375$ | 3 | 0.972 | 0.975 | 1.1505 | −2.4545 |
| | | $0.375 \le F < 0.5$ | 4 | 0.962 | 0.980 | 0.9074 | 0.4653 |
| | | $0.5 \le F < 0.625$ | 5 | 0.895 | 0.943 | 0.8210 | 2.6866 |
| | | $0.625 \le F < 0.75$ | 6 | 0.968 | 0.962 | 0.9297 | 0.4328 |
| | | $0.75 \le F < 0.875$ | 7 | 0.969 | 0.979 | 0.8442 | 0.9568 |
| | | $0.875 \le F < 1$ | 8 | 0.959 | 0.971 | 0.9537 | −0.7122 |
| Scale $a_{CNB}$ to $am$ | 2002–2009 | $0 \le F < 1$ | All | 0.969 | | 1.0994 | −0.0176 |
| Scale $a_{TOO}/s$ to $[aa_{HS}]_{CNB}$ | 1980 | $0 \le F < 0.125$ | 1 | 0.960 | 0.975 | 0.9630 | 1.6383 |
| | | $0.125 \le F < 0.25$ | 2 | 0.970 | 0.985 | 0.9625 | 0.7734 |
| | | $0.25 \le F < 0.375$ | 3 | 0.973 | 0.965 | 0.9236 | 1.7372 |
| | | $0.375 \le F < 0.5$ | 4 | 0.954 | 0.961 | 0.9844 | −0.4578 |
| | | $0.5 \le F < 0.625$ | 5 | 0.975 | 0.968 | 0.8295 | 2.0492 |
| | | $0.625 \le F < 0.75$ | 6 | 0.974 | 0.973 | 0.8942 | 1.2822 |
| | | $0.75 \le F < 0.875$ | 7 | 0.970 | 0.973 | 0.9565 | 1.0986 |
| | | $0.875 \le F < 1$ | 8 | 0.964 | 0.971 | 0.9573 | 0.8425 |
| Scale $a_{MEL}/s$ to $[aa_{HS}]_{TOO}$ | 1920 | $0 \le F < 0.125$ | 1 | 0.923 | 0.933 | 0.8934 | 0.8032 |
| | | $0.125 \le F < 0.25$ | 2 | 0.928 | 0.949 | 0.8589 | 0.8115 |
| | | $0.25 \le F < 0.375$ | 3 | 0.909 | 0.963 | 0.7325 | 2.5553 |
| | | $0.375 \le F < 0.5$ | 4 | 0.912 | 0.915 | 0.8085 | 0.8432 |
| | | $0.5 \le F < 0.625$ | 5 | 0.945 | 0.968 | 0.9564 | −0.0702 |
| | | $0.625 \le F < 0.75$ | 6 | 0.908 | 0.950 | 0.8264 | 0.5843 |
| | | $0.75 \le F < 0.875$ | 7 | 0.915 | 0.959 | 0.7737 | 1.8538 |
| | | $0.875 \le F < 1$ | 8 | 0.928 | 0.905 | 0.9100 | 0.6631 |

The first group of values in Table 1 gives the $s_c$ and $c_c$ values in each $F$ bin for this join between the HAD and ABN data. We here ascribe these values to the centre of the respective $F$ bin and used PCHIP interpolation to get the value required for the $F$ of a given $[aa_{HN}]_{ABN}$ data point. The annual variations in both $s_c$ and $c_c$ are of quite small amplitude but are often not of a simple form. This is not surprising considering the variety of different factors that could be influencing the variations with $F$, and that they are not generally the same at the two stations being inter-calibrated nor at Niemegk.

We use the variation with $F$ of both the scaling factor, $s_c$, and the offset, $c_c$, because at least some of the variation of the intercalibration with $F$ will be associated with the seasonal variation in the regular diurnal variations at the two sites and the background subtraction, which could give offset as well as gain (sensitivity) differences between the two sites.

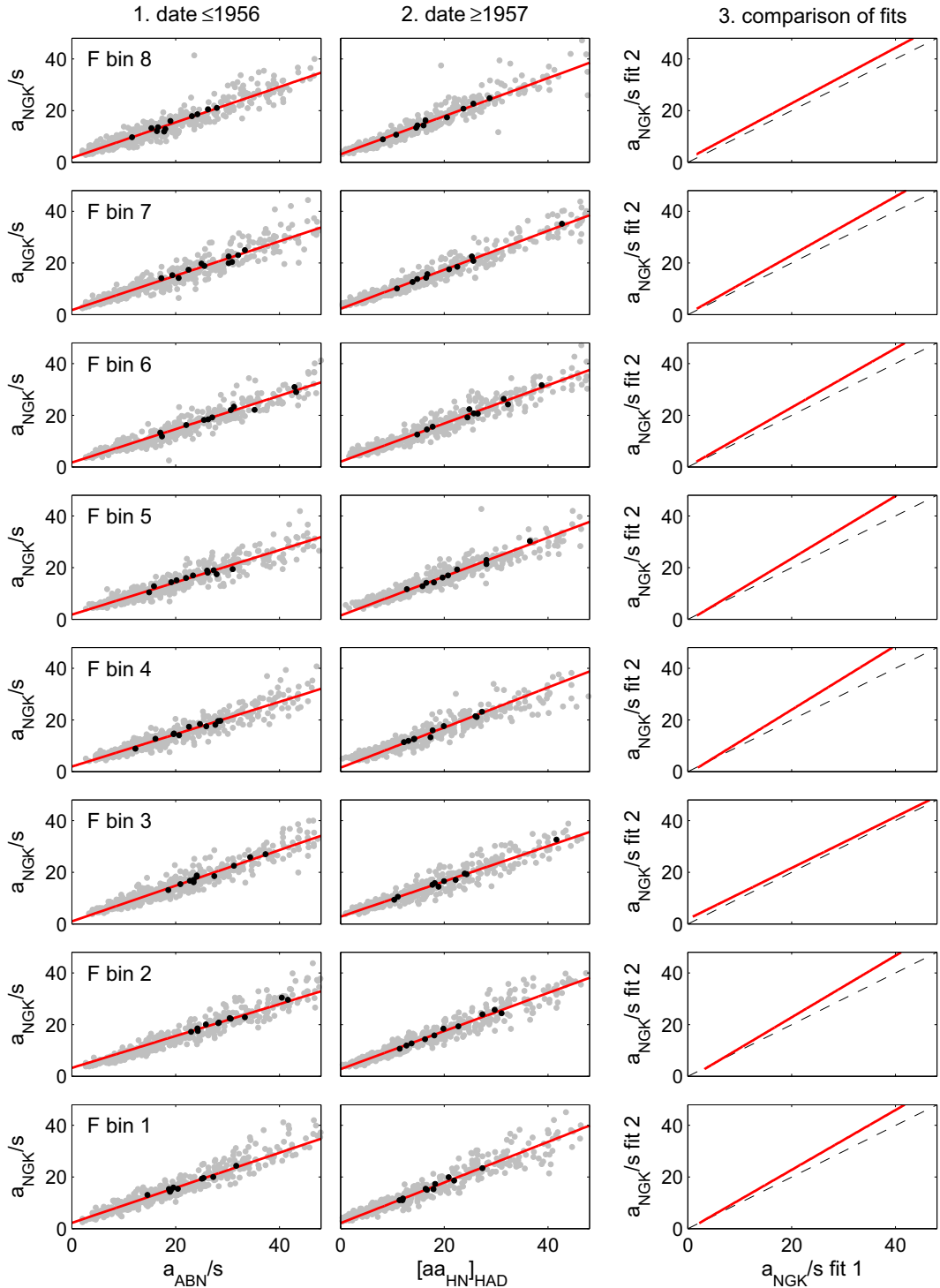

**Fig. 8.** The intercalibration of $aa_N$ data across the join between the Hartland (HAD) and Abinger (ABN) observations in 1957. The data are divided into eight equal-length fraction-of-year ($F$) bins, shown in the 8 rows, with the bottom row being bin 1 ($0 \leq F < 0.125$) and the top row being bin 8 ($0.875 \leq F < 1$). The left-hand column is for an interval of duration 11-years (approximately a solar cycle) before the join and shows scatter plots of the $aa$ data from Abinger (after division by $s(\delta)$ to allow for secular changes in the geomagnetic field) against the similarly-corrected simultaneous NGK data, $a_{NGK}/s(\delta)$. The middle column is for an interval of duration 11-years after the join and shows the corresponding relationship between the already-homogenized $aa$ data from Hartland $[aa_H]_{HAD}$ and the simultaneous $a_{NGK}/s(\delta)$ data. All axes are in units of nT. The grey dots are daily means to which a linear regression gives the red lines which are then checked against the annual means (for the $F$ bin in question) shown by the black dots. The right-hand column shows the fitted lines for the "before" interval, 1, against the corresponding fitted line for the "after" interval, 2: the red line would lie on the dotted line if the two stations had identical responses at the $F$ in question. The slope and intercept of these lines, giving the intercalibration of the two stations at that $F$, are given in Table 1.

Inspection of Figure 8 and Table 1 show that there is a variation with $F$ in the relationship between the two sites and our procedure takes account of this. Note that the intercept values are all small and that the red lines are actually shifted from the diagonal by the ratio of the classic $aa$ scalefactors. This emphasizes that the data from these two stations is, after allowance had been made for the secular geomagnetic drift through the $s(\delta)$ factor, similar. This reinforces the point that the large "calibration skip" between the Hartland and Abinger $aa_N$ values that has been widely discussed in the literature was, in the main, a necessary correction step to allow for the effects of the secular changes in the intrinsic field. Hence making a correction for this apparent calibration error, without first correcting for the temporal variation in the scaling factor $s(\delta)$, is only a first order correction and will give somewhat incorrect results in general.

As discussed above, we use Equation (2) to check the intercalibration data from Niemegk, where station XXX is SOD, LER and ESK for this join. If we do not correct for the effect of changing $\delta$ on the scaling factor $s(\delta)$ for these stations, we obtain values of $M_A/M_B$ of between 1.018 and 1.052, which implies there is drift in the average Neimegk data (to values that are slightly too low) of between about 3% and 5% over the intercalibration interval. However, after correcting the change in the stations' $\delta$ (in the same way as done for the $aa$ stations and Niemegk in Fig. 4) we get an $M_A/M_B$ of 1.053, 1.022 and 0.946 for LER, ESK and SOD, respectively. Giving these 3 estimates equal weight gives an average of 1.007, which implies the Niemegk calibration is stable to within 0.7% for our purposes. We note that this is not a test that we can repeat in such detail for all station joins. Hence we do not attempt to correct the NGK intercalibration data, beyond allowing for the effect of the drift in $\delta$ on $s(\delta)$. However, note that we will test this approach in the level of agreement in the final full $aa_{HN}$ and $aa_{HS}$ data sequences and in section 5, we will compare the long-term variation of these new $aa$ indices with the equivalent $IHV$ index as well as with $a_{NGK}/s(\delta)$, $a_{ESK}/s(\delta)$ and $a_{SOD}/s(\delta)$.

### 3.3 Inter-calibration of Abinger and Greenwich

Figure 9 corresponds to Figure 8, but is for the join between the Abinger and Greenwich data. Note that because the "after" data in this case are the corrected and re-scaled Abinger data, $[aa_{HN}]_{ABN}$ given by Equation (2), the slope and intercept values ($s_c$ and $c_c$) for this join are influenced by both the scaling of the Hartland data to $am$ and by the Abinger-to-Hartland join. Hence the calibration of Hartland against $am$ is passed back to Greenwich, as is in the nature of daisy-chaining. Given the data are taken from older generations of instruments and the fact that this second join is influenced by the first, we might have expected the plots to show more scatter than in Figure 8. In fact this is not the case and Table 1 shows the correlations are actually slightly higher for this intercalibration than the one discussed in the last section. Because concerns have been raised about a potential skip in the calibration of the $a_{NGK}$ composite in 1932, we use an "after" interval of 1926–1931 (inclusive, i.e. 6 years rather than the 11 years used for other joins). The correlations for all 8 $F$ bins were indeed found to be marginally lower if the full 11 years (1926–1936) were used but the regression coefficients were hardly influenced at all.

The corrected Sodankylä $K$-indices give $M_A/M_B = 0.943$ for this join which could imply a 6% problem with the Niemegk spline. However, we note that Sodankylä gave a lower value than the average for the Abinger-Hartland calibration interval which is likely to be a consequence of its close proximity to the auroral oval. As for that join, we here use the Niemegk data as a calibration spline without correction, but will test the result in Section 5.

The Greenwich data are intercalibrated using the equivalent equation to Equation (4):

$$[aa_{HN}]_{GRW}(F) = s_c(F) \cdot a_{GRW}(F)/s(\delta) + c_c(F) \qquad (6)$$

using the appropriate $s_c$ and $c_c$ values given in Table 1 and the interpolation in $F$ scheme described above.

### 3.4. Inter-calibration of the southern hemisphere stations

Figures 10 and 11 are the same as for Figure 8 for the joins between, respectively, the Canberra and Toolangi stations and between the Toolangi and Melbourne stations (note that the colours of the regression lines matches the colours used to define the joins in Fig. 2). The Toolangi and Melbourne data are corrected using the corresponding Equations to (4) and (5) to give $[aa_{HS}]_{TOO}$ and $[aa_{HS}]_{MEL}$.

Figure 10 uses the $am$ data to make the Canberra-Toolangi intercalibration but, as mentioned above, almost identical results were obtained if either the $as$ index or $a_{NGK}/s$ was used. Using $a_{NGK}/s$ did increase the scatter in the daily values slightly, but the regression fits remained almost exactly the same. In the case of the Toolangi-Melbourne join, the best comparison data available are the Niemegk $K$ indices, but based on the above experience of using it for the Canberra-Toolangi join, it is not a major concern that the intercalibration data are from the opposite hemisphere, although, as expected, it does increase the scatter between the daily values.

Note that the only operation to make $aa_{HN}$ and $aa_{HS}$ similar is the scaling of both to $am$ over the interval 2002–2009, achieved by Equations (2) and (3). Thereafter the northern and southern data series are generated independently of each other. Therefore the degree to which the two hemispheric indices agree with each other over time becomes a test of the intercalibrations and the stability of the datasets.

## 4 The homogeneous composite

We can then put together 150-year composite of $aa_{HN}$ (using $[aa_{HN}]_{GRW}$ $[aa_{HN}]_{ABN}$, and $[aa_{HN}]_{HAD}$) and the red line in Figure 2b shows the resulting variations in annual means. The blue line is the corresponding composite of $aa_{HS}$ (using $[aa_{HS}]_{MEL}$, $[aa_{HS}]_{TOO}$, and $[aa_{HS}]_{CNB}$). Comparison with Figure 2a shows that the calibrations described in the previous section have produced hemispheric data series which agree much more closely with each other than do $aa_N$ and $aa_S$. To quantify the improvement, Figure 12 compares the distributions of the differences in daily means of northern and southern hemisphere indices in 50-year intervals, $\Delta_{NS}$. The top row is for the classic $aa$ indices (so $\Delta_{NS} = aa_N - aa_S$). The bottom row is for the homogenised $aa$ indices (so $\Delta_{NS} = aa_{HN} - aa_{HS}$). The left column is for 1868–1917 (inclusive); the middle column for 1918–1967; and the right-hand column for 1968–2017. Note that distributions are narrower and taller for the first time interval because mean values were lower and so hemispheric differences are correspondingly lower.

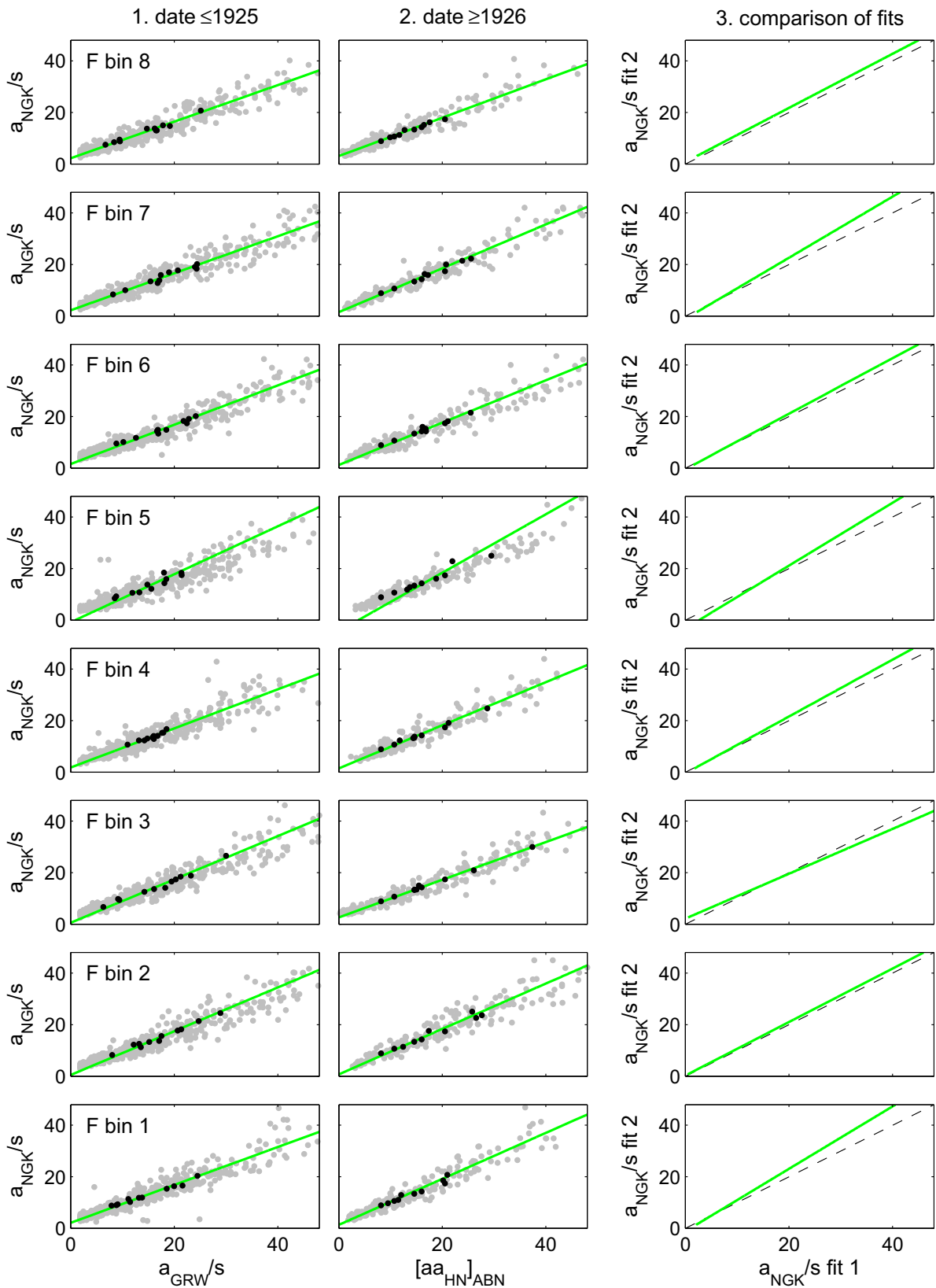

**Fig. 9.** The same as Figure 8 for the join between the Greenwich and Abinger data.

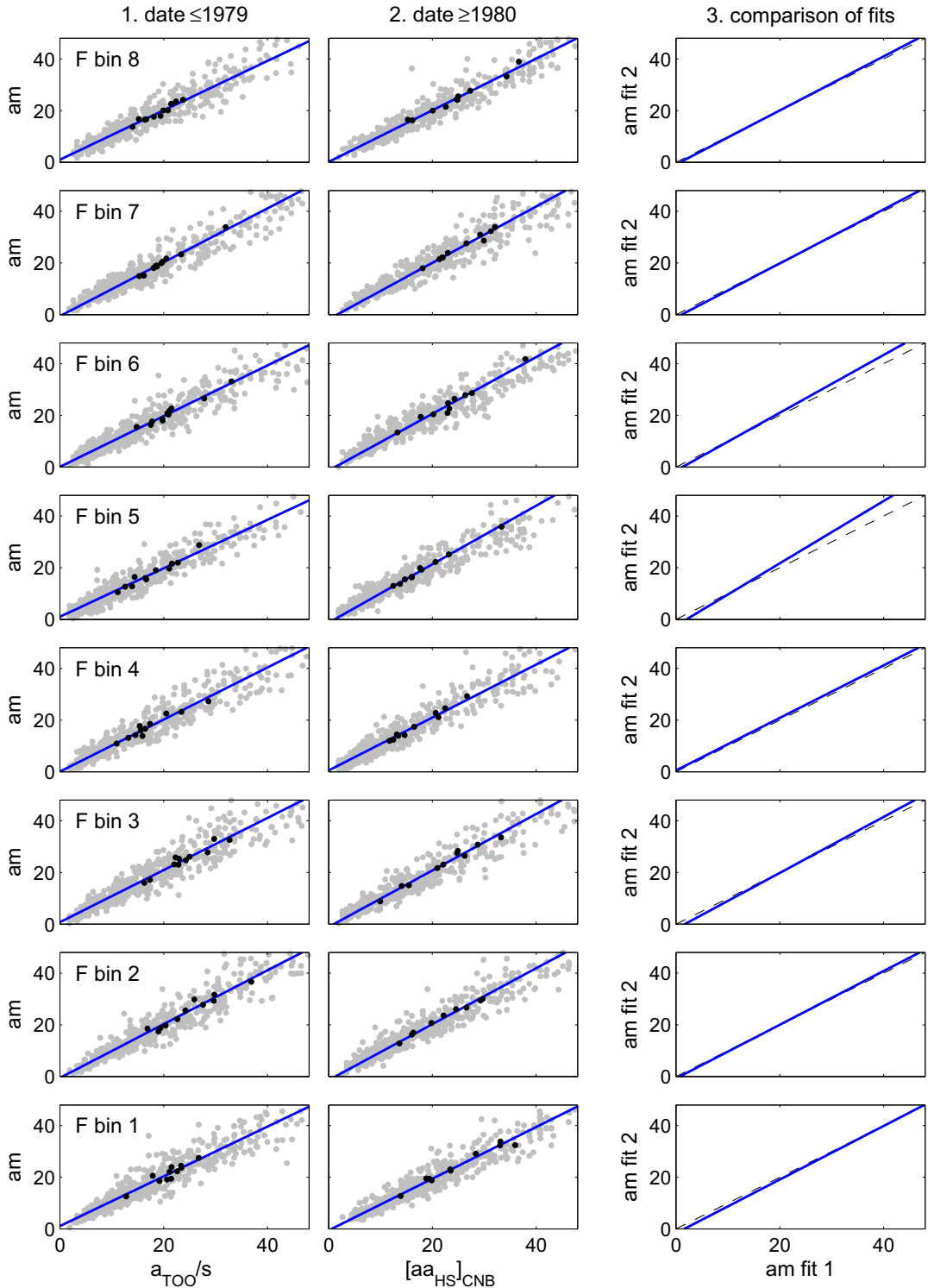

**Fig. 10.** The same as Figure 8 for the join between the Toolangi and Canberra data.

A number of improvements can be seen in the distributions for $(aa_{HN} - aa_{HS})$, compared to those for $(aa_N - aa_S)$. Firstly the mean of the distributions has been reduced to zero (to within $10^{-3}$) in all three time intervals by the homogenized index. Not only is this smaller than for the corresponding classical index, but also the upward drift in the mean value $\Delta_{NS}$ has been removed. This improvement in the mean difference quantifies the improvement that can be seen visually by comparing Figures 2a and b. Secondly, the width of the distribution in ($aa_{HN} - aa_{HS}$) is always lower than for the corresponding distribution of ($aa_N - aa_S$): this can be seen in the given values of the standard deviation, $\sigma_{\Delta NS}$ and in the separation of the decile values (which are given by the vertical green lines). Thirdly the $\Delta_{NS}$ distributions for the classic index show a marked asymmetry: this can be seen by the fact that the median of the distributions (vertical cyan line) is consistently smaller than the mean and that the modulus of the lower decile value is always less than the upper decile value. This asymmetry has been removed completely in the homogenized data series after 1917. (For 1868–1917 the 1-$\sigma$ points are symmetrical but the mode is slightly lower than the mean.) Lastly the distributions for the classic index show a tendency for quantized levels (particularly for 1868–1917) and more kurtosis in shape than for the homogenized indices. On the other hand, ($aa_{HN} - aa_{HS}$) shows very close to a Gaussian form at all times. If there is a physical reason why the distribution should diverge from a Gaussian, it is not clear. Hence, agreement between the northern and southern hemisphere indices has been improved, in many aspects, by the process described in this paper.

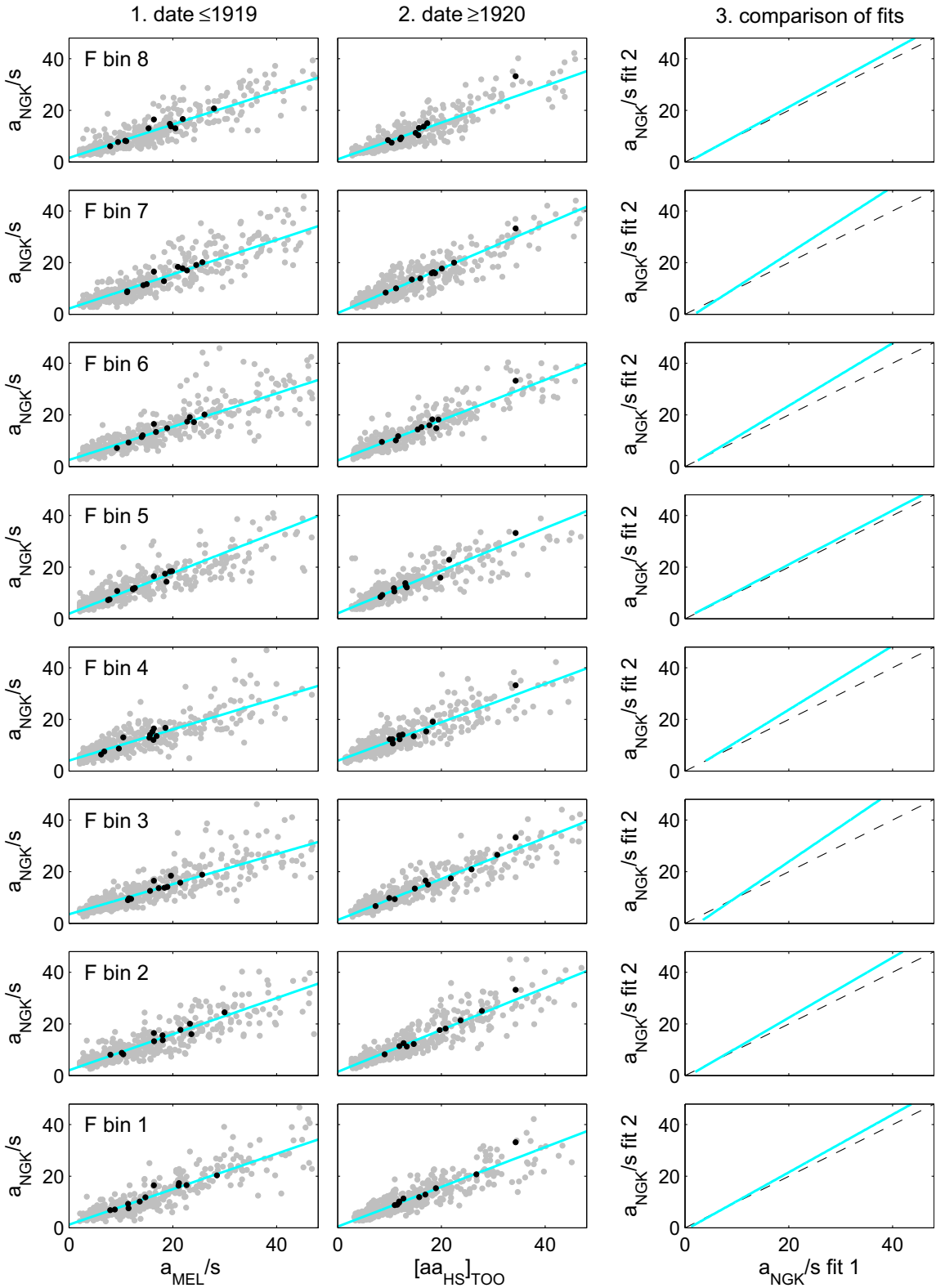

**Fig. 11.** The same as Figure 8 for the join between the Melbourne and Toolangi data.

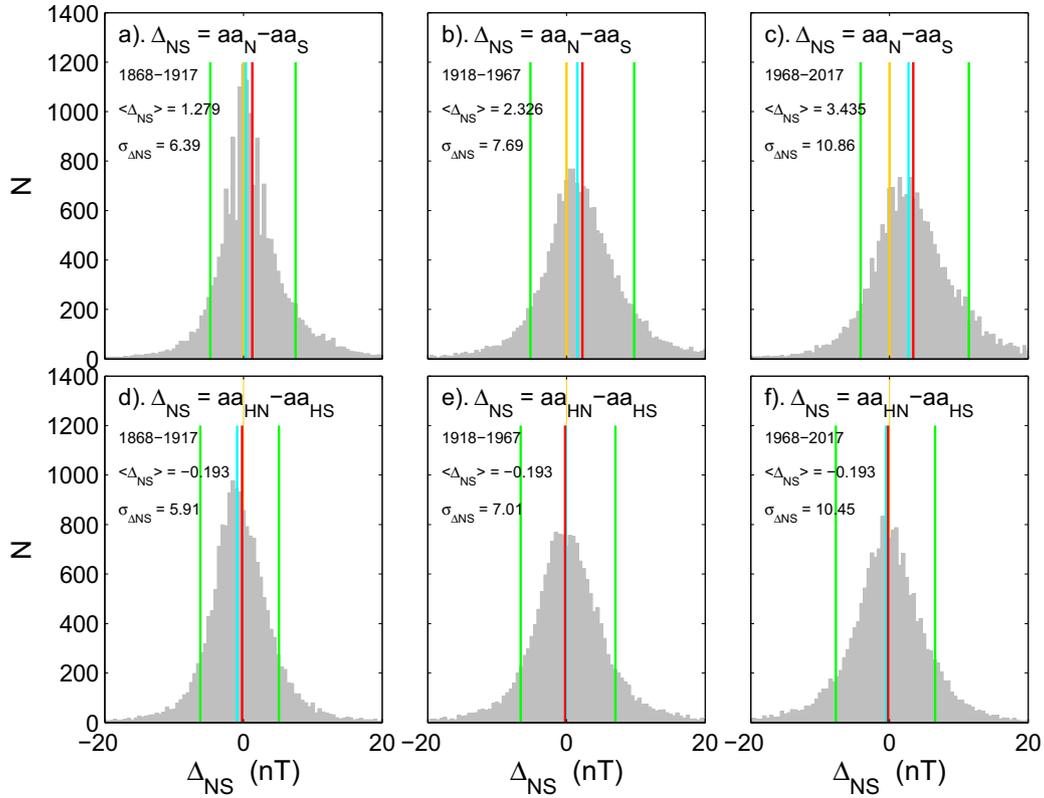

**Fig. 12.** Distributions of the differences in daily means of northern and southern hemisphere indices, $\Delta_{NS}$, for 50-year intervals. The top row is for the classic $aa$ indices, so that $\Delta_{NS} = aa_N - aa_S$. The bottom row is for the homogenised $aa$ indices, so that $\Delta_{NS} = aa_{HN} - aa_{HS}$. Parts (a) and (d) are for 1868–1917 (inclusive); parts (b) and (e) are for 1918–1967; and parts (c) and (f) are for 1968–2017. In each panel, the vertical orange line is at $\Delta_{NS} = 0$, the vertical cyan line is the median of the distribution, the vertical red line the mean ($\langle\Delta_{NS}\rangle$), and the green lines the upper and lower deciles. Note they are plotted in the order, orange, cyan, then red and so the mean can overplot the others (this particularly occurs in the bottom row). In each panel the distribution mean, $\langle\Delta_{NS}\rangle$ and the standard deviation, $\sigma_{\Delta NS}$, are given.

Lastly, Figure 2c compares the annual means of the homogenised $aa$ index derived here, defined by

$$aa_H = (aa_{HN} + aa_{HS})/2 \qquad (7)$$

with the classic $aa$ index and the corrected $aa$ index, $aa_C$, that was generated by Lockwood et al. (2014) by correcting the classic $aa$ index for the Hartland-Abinger intercalibration using the $Ap$ index. The black line is the $aa_H$ index from Equation (6) and so contains allowance for the secular drift in the main field and for the re-calibration of stations presented in Section 3. The mauve line is the classic $aa$ index. It can be seen that, because of the scaling to the recent $am$ index data, the $aa_H$ index values are always a bit lower than $aa$. The cyan line and points show annual means in the $am$ index. It is noticeable that as we go back in time towards the start of these data, these $am$ means follow the classic $aa$ index rather well and so become slightly larger than the corresponding annual means in $aa_H$. This indicates that the secular drift in the intrinsic geomagnetic field is having an influence on even $am$ over its lifetime. The orange line is the corrected $aa$ data series, $aa_C$. By definition, this is the same as $aa$ before the Abinger-Hartland join 1957: hence the orange line lies underneath the mauve one in this interval. Between 1957 and 1981, $aa_C$ is slightly larger than $aa_H$ most of the time, but after 1981 the orange line can no longer be seen because it is so similar to $aa_H$. Hence correcting for the Abinger-Hartland join, without correcting for the effects of the secular drift in the intrinsic field have caused corrected indices such as $aa_C$ (and others like it) to underestimate the upward rise in $aa_H$.

Taking 11-point running means to average out the solar cycle, $aa$, $aa_C$ and $aa_H$ all give smoothed minima in 1902 of, respectively, 11.66 nT, 11.77 nT and 10.87 nT. The maxima for $aa$ and $aa_H$ are both in 1987, shortly after the peak of the sunspot grand maximum (Lockwood & Fröhlich, 2007), being 27.03 nT and 24.25 nT, giving a rises of 15.37 nT in $aa$ and 13.38 nT in $aa_H$ over the interval 1902–1987. The corrected index, $aa_C$, is somewhat different with a value of 24.51 nT in 1987, but a slightly larger peak of 24.80 nT in 1955. Over the interval 1902–1987 the rise in $aa_C$ is 12.73 nT.

# 5 Comparison of the homogenized $aa$ index with the $IHV$ index and corrected $a_k$ values from Niemegk, Eskdalemuir and Sodankylä

The development of the Inter-Hour Variability ($IHV$) index was discussed in Section 1.2. The most recent version was published by Svalgaard & Cliver (2007). It is based on hourly means of the observed horizontal magnetic field at each station and its compilation is considerably simpler than, and completely different to, that of the range indices such as $aa$. It is defined as the sum of the unsigned differences between adjacent hourly means over a 7-hour interval centered on local midnight (in solar local time, not magnetic local time). The daytime hours are excluded to reduce the effect of the regular diurnal variation and UT variations are removed assuming an equinoctial time-of-day/time-of-year pattern, which reduces the requirement to have a network of stations with full longitudinal coverage. Using data from 1996–2003, Svalgaard & Cliver (2007) showed that $IHV$ has major peaks in the auroral ovals, but equatorward of $|\Lambda_{CG}| = 55°$ it could be normalized to the latitude of Niemegk using a simple ad-hoc function of $\Lambda_{CG}$. Note that $IHV$ does not allow for the changes in the stations' $\Lambda_{CG}$ due to the secular change in geomagnetic field. This will be a smaller factor for $IHV$ than for the range indices as the latitude dependence is weaker. However, in $IHV$ this effect will also be convolved with that of the changing distribution and number of available stations. This is because the number of stations contributing to the annual mean $IHV$ values tabulated by Svalgaard & Cliver (2007) varies, with just one for 1883–1889, two for 1889–1900, rising to 51 in 1979 and before falling again to 47 in 2003. Although the removal of the diurnal variation (by assuming an equinoctial variation) and the removal of the $\Lambda_{CG}$ variation (by using the polynomial fit to the latitudinal variation in the 1996–2003 data) allows the $IHV$ index to be compiled even if only one station is available, such an index value will have a much greater uncertainty because it will not have the noise suppression that is achieved by averaging the results from many stations in later years. It must be remembered, therefore, that the uncertainties in the $IHV$ index increase as we go back in time.

Lockwood et al. (2014) show that in annual mean data $IHV$ correlates well (correlation coefficient, $r = 0.952$) with $BV_{SW}{}^n$, where $B$ is the IMF field strength, $V_{SW}$ is the solar wind speed and $n = 1.6 \pm 0.8$ (the uncertainty being at the 1-$\sigma$ level), whereas the corrected $aa$ index and gave $r = 0.961$ with $n = 1.7 \pm 0.8$. The difference in the exponent $n$ is small (and not statistically significant) and so we would expect the long-term and solar cycle variations in $aa$ and $IHV$ to be very well correlated. Indeed, Svalgaard & Cliver (2007) found that even in Bartel's rotation period (27-day) means $IHV$ and the range $am$ index were highly correlated ($r = 0.979$).

Figures 13a–f compare annual means of the new homogenised indices $aa_H$, $aa_{HN}$ and $aa_{HS}$ with the $IHV$ index. The left-hand plots show the time series and the best-fit linear regression of $IHV$. The right plots so scatter plots of the new indices against $IHV$ and the least squares best-fit linear regression line in each case. For comparison, the bottom panel compares the hemispheric homogenized indices $aa_{HN}$ and $aa_{HS}$. The agreement is extremely good in all cases: for $aa_{HS}$ and $IHV$ the coefficient of determination is $r^2 = 0.937$; for $aa_{HN}$ and $IHV$ $r^2 = 0.962$; for $aa_H$ and $IHV$, $r^2 = 0.958$; and for $aa_{HN}$ and $aa_{HS}$, $r^2 = 0.992$. This level of agreement is exceptionally high, considering $IHV$ is constructed in an entirely different manner, and from different data and with different assumptions (e.g. it assumes an equinoctial time-of-day/time-of-year pattern). In particular, note that $IHV$ is not homogeneous in its construction as the number of stations contributing decreases as we go back in time: this would increase random noise but not explain systematic differences. Also $IHV$ only uses nightside data whereas $k$ indices use data from all local times; however, $k$ indices respond primarily to substorms (see supplementary material file of Lockwood et al., 2018c) which occur in the midnight sector. Also shown in Figure 13 are the corresponding comparisons with the corrected $a_K$ indices from Niemegk, Sodankylä, and Eskdalemuir $a_{NGK}/s(\delta)$, $a_{SOD}/s(\delta)$ and $a_{ESK}/s(\delta)$ (parts g/h, i/j, and k/l respectively). The coefficients of determination ($r^2$) are 0.945 and 0.958 and 0.914, respectively.

Hence Figure 13 is a good test of the intercalibrations used in constructing $aa_{HN}$ and $aa_{HS}$ in the context of annual mean data. There are differences between all the regressed variations but they are small. The internal correlation between the hemispheric $aa$ indices is now greater than that with any other equivalent data series: the worst disagreements are that $aa_{HN}$ exceeds $aa_{HS}$ around the peak of solar cycle 17 (around 1940) and $aa_{HS}$ exceeds $aa_{HN}$ around the peak of solar cycle 14 (around 1907). In both cases, the independent data in Figure 13 indicate that the error is in both $aa_{HN}$ and $aa_{HS}$ as these data follow $aa_H$ more closely. In the case of the largest error (around 1940), $IHV$, $a_{NGK}/s(\delta)$, $a_{SOD}/s(\delta)$ and $a_{ESK}/s(\delta)$ all also suggest that $aa_{HS}$ is an underestimate by slightly more than $aa_{HN}$ is an overestimate and so $aa_H$ is very slightly underestimated, but only by less than 0.5 nT. It should be noted that this largest deviation between $aa_{HN}$ and $aa_{HS}$ occurs when the data are supplied by the Abinger and Toolangi observatories, respectively and that Figure 2b shows that $aa_{HN}$ and $aa_{HS}$ agree more closely both earlier and later in the interval 1925–1956 when these two stations are used. Hence the deviation is caused by relative drifts in the data from these stations and not by the inter-calibrations developed in this paper.

The grey areas in the left-hand panels of Figure 13 show the estimated $\pm 1\sigma$ uncertainty in annual $aa_H$ estimates, where $\sigma = 0.86$ nT is the standard deviation of the distribution of annual ($aa_{HN} - aa_{HS}$) values.

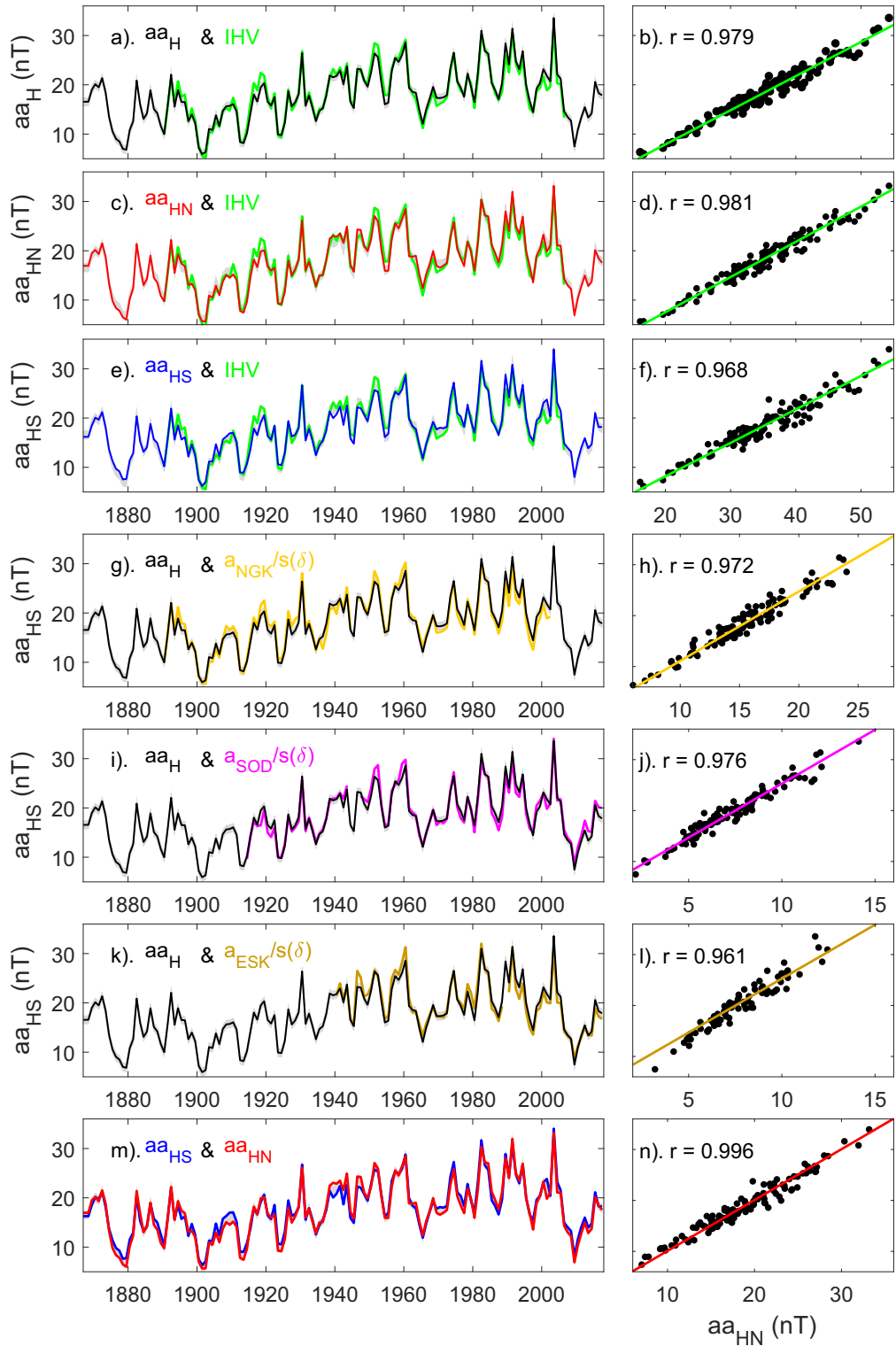

**Fig. 13.** Comparison of the new homogenised indices $aa_H$, $aa_{HN}$ and $aa_{HS}$, with the *IHV* index and the $\delta$-corrected $a_{NGK}$, $a_{SOD}$ and $a_{ESK}$ values. The left-hand plots show the time series and their respective best-fit linear regressions. The right-hand plots show scatter plots of the new indices against the test indices ($a_{NGK}/s(\delta)$, $a_{SOD}/s(\delta)$ or *IHV*) and the least-squares best-fit linear regression line. The linear correlation coefficient *r* is given in each case. Parts (a) and (b) are for $aa_{HN}$ and *IHV*; parts (c) and (d) are for $aa_{HS}$ and *IHV*; parts (e) and (f) are for $aa_{HS}$ and *IHV*; parts (g) and (h) are for $aa_H$ and the corrected $a_K$ values from Niemegk, $a_{NGK}/s(\delta)$; parts (i) and (j) are for $aa_H$ and the $\delta$-corrected $a_K$ values from Sodankylä, $a_{NGK}/s(\delta)$, parts (k) and (l) are for $aa_H$ and the $\delta$-corrected $a_K$ values from Eskdalemuir, $a_{ESK}/s(\delta)$. For comparison, the bottom panel (parts m and n) compares the hemispheric homogenized indices $aa_{HN}$ and $aa_{HS}$. In each panel, the grey area defines the estimated ±1σ uncertainty in $aa_H$.

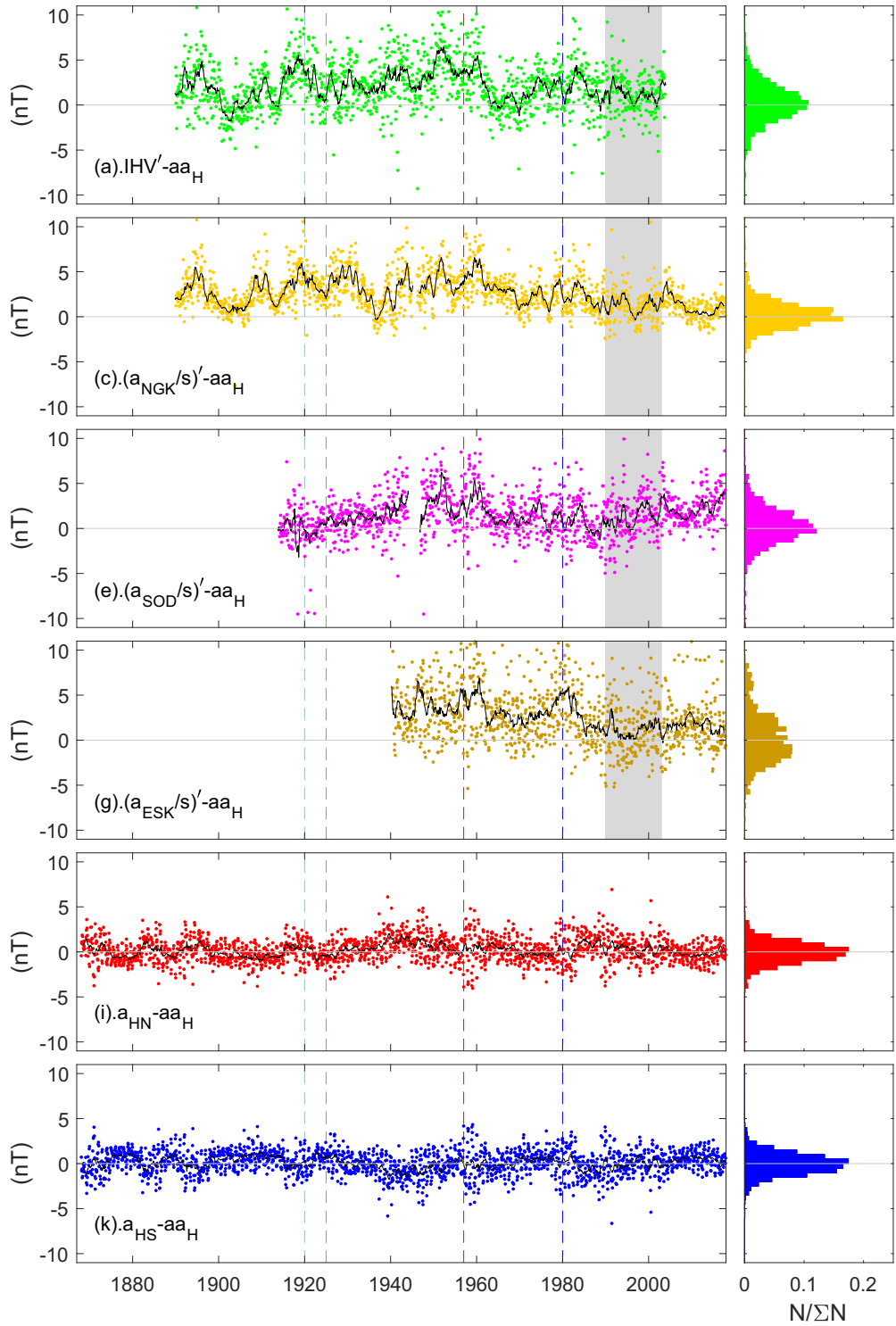

**Fig. 14.** (Left-hand plots) Bartels rotation interval (27-day) means of the deviation of various scaled indices from the new $aa_H$ index (points) and 13-point running means of those 27-day means (black lines). The indices are all scaled by linear regression to the $aa_H$ index over the interval 1990–2003 (the interval shaded in gray). A prime on a parameter denotes that this scaling has been carried out. (Right-hand plots) The distribution of the deviations of the 27-day means (the dots in the corresponding left-hand plot) from their simultaneous 13-point smoothed running means (the black line in the corresponding left-hand plot). (a) and (b) are for the scaled IHV index, $IHV'$; (c) and (d) for the corrected and scaled $a_K$ index from Niemegk, $(a_{NGK}/s)'$; (e) and (f) for the corrected and scaled $a_K$ index from Sodankylä, $(a_{SOD}/s)'$; (g) and (h) for the corrected and scaled $a_K$ index from Eskdalemuir, $(a_{ESK}/s)'$; (i) and (j) for the homogenized northern hemisphere $aa$ index, $aa_{HN}$; and (k) and (l) for the homogenized southern hemisphere $aa$ index, $aa_{HS}$. The dashed lines in the left-hand plots mark the dates of $aa$ station joins, using the same colour scheme as used in Figure 2. Horizontal grey lines are at zero deviation.

Figure 14 gives a more stringent test of the station joins in the $aa_H$ index using means over 27-day Bartels rotation intervals. The dots in the left-hand panels show the deviations of 27-means of the $aa_H$ index from scaled test indices. Long-term trends in those deviations are searched for by looking at 13-point running means (just under 1 year) of those deviations, shown by the black lines. The histograms to the right give the overall occurrence distribution of the deviation of the dots from the blacked lines and hence give an indication of the scatter around the *trend*. In all cases, the test data are well correlated with $aa_H$, the linear correlation coefficients for Bartels rotation means being: 0.938 for $IHV$; 0.964 for $a_{NGK}/s$; 0.948 for $a_{SOD}/s$; and 0.900 for $a_{ESK}/s$. For comparison, the values are 0.989 and 0.987 for $aa_{HN}$ and $aa_{HS}$ (but of course they are not independent data from $aa_H$). To derive the deviation, the test index is linearly regressed with $aa_H$ index over a common calibration interval of 1990–2003 (for which data are available for all indices) and test data that have been scaled $aa_H$ using the least-squares linear regression fit for this interval are denoted with a prime. What we are searching for are consistent step-like change in the deviations (on timescales comparable to the ~1 year smoothing time constant) at the time of one of the $aa$ station joins (which are marked by the vertical dashed lines using the same colour scheme as in Fig. 2). The appearance of such a step in several of the test data series would indicate a calibration error in $aa_H$. The most-tested join is that between Toolangi and Canberra in $aa_{HS}$ (blue dashed line). In generating $aa_H$ this was calibrated using a spline of the $am$ index. The only step-like feature is shortly after that join in the Eskdalemuir data and, as this is the least well correlated of the test indices and shows the latest scatter in its 27-day means, this is not good evidence for a problem with this join. The much-discussed Abinger-to-Hartland join (red dashed line) also does not produce a consistent signature in the test data, with equal mean deviations before and after in all the corrected $a_K$-index data, i.e. from Sodankyla, Neimegk, Eskdalemuir and $aa_{HS}$ (which at this time is data from Toolangi). There is a small step in the Niemegk deviations around 1970, but this is after the calibration interval for this join (which is 1946–1967). The deviations for $IHV$ show a step between 1960.5 and 1963.5, but this is after the join and coincides with the large fall in all indices seen at the end of solar cycle 19. Hence this appears to be related to a slight non-linearity between the $IHV$ index and range-based indices, rather than the calibration join. For the two earlier joins, Greenwich-to-Abinger (green dashed line) and Melbourne-to-Toolangi (cyan dashed line), the independent test data available are $IHV$, the $aa_H$ data from the opposite hemisphere and to a lesser extent $a_{SOD}/s$ (which only extends back to 1914 which is only 6 years before the MEL-TOO join). Note, however, that $IHV$ is compiled using hourly means from just 2 stations before 1900, rising to 11 by 1920, and one of stations is Niemegk, and so it does not provide a fully independent test of our station inter-calibrations. We note that the data use to make the joins, $a_{NGK}/s$, show some fluctuation over the relevant intervals, but no consistent step. The $IHV$ data do show a step 3 years before the MEL-TOO join but this is at the same time as the strong rise in all indices at the start of cycle 15 from the very low values during the minimum between cycles 14 and 15. Hence, as for the 1961/2 step in $IHV$, this appears to be more associated with a slight non-linearity between $IHV$ and $a_K/s$ values at low activity than with a calibration skip caused by an $aa$ station change.

# 6 Conclusions

The classic $aa$ indices now cover 150 years, an interval long enough that there are significant effects on the indices due to the effects of secular changes in the intrinsic geomagnetic field. We here correct for these using the standard approach to calculating $K$-indices, but making the scale factors employed for each station a function of time. We also show that this improves the inter-calibration of the range-based data from other stations which had also been influenced by the assumption of constant scale factors. The intercalibrations are shown, in general, to depend on time-of year ($F$), which is here accommodated using 8 equal-sized bins in $F$ and interpolating to the date of each 3-hourly measurement. This allows us to correct for seasonal effects on both the instrumentation and the background subtraction procedures.

We call the corrected data series that we have produced the "homogenized" $aa$ data series, because it eliminates a number of differences between the data series from the two hemispheres. In this paper we have concentrated on the results in annual means. In a companion paper (Paper 2), we make further allowances for the effects of the variations of each station's sensitivity with time-of-day and time-of-year (Lockwood et al., 2018b). These further corrections are carried out such that the annual means presented here ($\langle aa_H \rangle_{\tau=1yr}$, $\langle aa_{HN} \rangle_{\tau=1yr}$ and $\langle aa_{HS} \rangle_{\tau=1yr}$) remain unchanged. In the supplementary material to the present paper we give the annual mean values of the homogenized $aa_H$, $aa_{HN}$, and $aa_{HS}$ data series as these will be subject to no further corrections. We also attach a file containing the annual $\delta$ and $s(\delta)$ values used to make allowance for the secular field changes. The supplementary data attached to Paper 2 will give the daily and 3-hourly values of $aa_H$, $aa_{HN}$ and $aa_{HS}$. The equations given in the text of the present paper, along with the coefficients given in Table 1, give a complete recipe for generating this first level of the $aa$ homogenized data series from the classical $aa$ values.

Given the close agreement between the independently-calibrated $aa_{HN}$ and $aa_{HS}$ indices, we can be confident that the 13.38 nT rise in the 11-year averages of $aa_H$ is accurate. The standard error in the difference of annual means of $aa_{HN}$ and $aa_{HS}$ is $\xi_1 = 0.082$ nT for 1996–2017 and $\xi_2 = 0.039$ nT for 1868–1889. Treating these as the uncertainties in the average levels in these intervals gives an estimate uncertainty in the difference between them of $(\xi_1{}^2 + \xi_2{}^2)^{1/2} = 0.091$ nT which is just ≈1% of the 13.38 nT rise in $aa_H$. Also, because $aa_{HN}$ and $aa_{HS}$ are calibrated against the $am$ index over the last 5 years, we can also be confident that the values of 10.87 nT and 24.25 nT for 1902 and 1987 are correct to within the above accuracies. Thus the new 11-year smoothed $aa_H$ values reveal a rise of 123% between these two dates. In comparison, the corresponding rise in the classic $aa$ between these dates was 15.36 nT (132%) and that in $aa_C$ was 12.73 nT (108%). Therefore, although the rise in the classic $aa$ was excessive (by 9%), correcting for the Abinger-Hartland intercalibration in isolation, without allowing for the drifts caused by the secular change in the main field, gives an over-correction and the rise in $aa_C$ is here found to be too small by 15%.

As a result, reconstructions of solar wind parameters that have been based on $aa_C$, such as the open solar flux, the solar wind speed and the near-Earth IMF (Lockwood et al., 2014) will also have underestimated the rise that took place during the 20th century and may not have the correct temporal waveform (given that the peak $aa_C$ was in 1955 rather than 1987). However, it should be noted that $aa_C$ was one of four geomagnetic indices (including *IHV*) used by Lockwood et al. (2014) which means the effect of its underestimation of the rise will be reduced in the reconstructions. Replacing $aa_C$ with $aa_H$ should also have the effect of reducing the uncertainties. Note also that the dependencies of the various indices on the IMF, *B*, and the solar wind speed, $V_{SW}$, are such that it is the $V_{SW}$ estimates that will be most affected. This will be investigated, and the reconstructions amended, in subsequent publications.

A point of general importance to geomagnetic indices is that, as shown in Figure 2c, the *am* index follows the classic *aa* series very closely, but as we go back in time towards the start of the *am* index in 1959, the homogenized index $aa_H$ becomes progressively smaller than *aa* by a consistent amount. Similarly, close inspection of Figure 13 shows that *IHV* has fallen by slightly more than $aa_H$ in this interval. These differences are the effects of changes in the intrinsic geomagnetic field on the indices. In the case of all previous indices based on *K* values (*Kp*, *ap*, *am*, *an*, and the classic *aa*) this arises from drift in the *L* values (the threshold value of the *K* = 9 band) in the case of *IHV* it arises from the normalization to a reference latitude (and potentially also from the use of an equinoctial pattern used to remove the diurnal variation). Hence we conclude that all geomagnetic indices should make allowance for these effects if they are to be fit for purpose in studies of space climate change.

## Supplementary material

*Acknowledgements.* Without either group knowing it, much of the work reported in this paper was carried out completely independently at University of Reading (UoR) and at École et Observatoire des Sciences de la Terre (EOST) but when we discovered that we had independently reached almost identical results, we decided a single publication would minimise duplication and potential confusion. The authors thank all staff of ISGI and collaborating institutes, past and present, for the long-term compilation and databasing of the *aa* and *am* indices which were downloaded from http://isgi.unistra.fr/data_download.php. We also thank the staff of Geoscience Australia, Canberra for the southern hemisphere *aa*-station *K*-index data, and colleagues at British Geological Survey (BGS), Edinburgh for the northern hemisphere *aa*-station *K*-index data, and the Helmholtz Centre Potsdam of GFZ, the German Research Centre for Geosciences for the Niemegk *K*-index data. The work at UoR and BGS is supported by the SWIGS NERC Directed Highlight Topic Grant number NE/P016928/1 with some additional support at UoR from STFC consolidated grant number ST/M000885/1. The work at EOST is supported by CNES, France. The editor thanks two anonymous referees for their assistance in evaluating this paper.